\documentclass[aps,superscriptaddress,
twocolumn,
pre,floatfix,showpacs,footnote]{revtex4-1}
\usepackage{epsf,color,colordvi}
\usepackage{graphicx}
\usepackage{amsmath,amssymb}
\usepackage{mathtools}
\usepackage{hyperref}
\usepackage{times}
\usepackage[colorinlistoftodos,prependcaption,textsize=tiny]{todonotes}
\usepackage{pstricks}
\usepackage{placeins}
\usepackage{ulem} 
\usepackage[mathscr]{euscript}
\usepackage{comment}

\newcommand{\ts}{\textstyle }

\newcommand{\etal}{\textit{et al.}}

\newcommand{\hH}{\widehat{H}}
\newcommand{\hK}{\widehat{K}}
\newcommand{\hV}{\widehat{V}}
\newcommand{\cJ}{\mathscr{J}}
\newcommand{\cM}{\mathscr{M}}
\newcommand{\cT}{\mathscr{T}}
\newcommand{\hTheta}{\widehat{\Theta}}

\begin{document}
\title{Numerical solutions of the time-dependent Schr\"odinger equation in two dimensions}

\author{Wytse van Dijk}
\affiliation{Department of Physics, Redeemer University College, Ancaster, Ontario L9K 1J4, Canada}
\affiliation{Department of Physics and Astronomy,
McMaster University, Hamilton, Ontario L8S 4M1, Canada}
\email{vandijk@physics.mcmaster.ca} 

\author{Trevor Vanderwoerd}
\affiliation{Department of Physics, Redeemer University College, Ancaster, Ontario L9K 1J4, Canada}

\author{Sjirk-Jan Prins}
\affiliation{Department of Physics, Redeemer University College, Ancaster, Ontario L9K 1J4, Canada}

\date{\today}


\begin{abstract}
The generalized Crank-Nicolson method is employed to obtain numerical solutions of the two-dimensional time-dependent Schr\"odinger equation.  An adapted alternating-direction implicit method is used, along with a high-order finite difference scheme in space.  Extra care has to be taken for the needed precision of the time development. The method permits a systematic study of the accuracy and efficiency in terms of powers of the spatial and temporal step sizes.  To illustrate its utility the method is applied to several two-dimensional systems.
\end{abstract}


\maketitle 

\section{Introduction}

The determination of accurate numerical solutions of the time-dependent Schr\"odinger equation is an ongoing enterprise.  The quantum wave equation is fundamental to the understanding of nonrelativistic atomic and subatomic systems and phenomena.  Consequently it occurs in a diversity of physical systems.  Ideally analytic solutions are available, but most realistic situations are too complex to yield such solutions. 

In the last few years a number of improvements have been made to yield more accurate solutions with greater efficiency.  The type of method often depends on the problem at hand, i.e., dimensionality, time dependence of the interaction, short- or long-time behaviour, etc.   The ``method of choice" for some years is the Chebyshev polynomial expansion of the time-evolution operator with (inverse) Fourier transformations to deal with the spatial development as time progresses~\cite{talezer84,leforestier91}.  More recently the Pad\'e approximant representation of the time-evolution operator is exploited~\cite{puzynin99,*puzynin00,vandijk07,vandijk11,vandijk14a,vandijk16}.  This approach is unitary, stable, and allows for systematic estimate of errors in terms of powers of the temporal and spatial step sizes.  The two approaches have been shown to have comparable efficacy~\cite{formanek10,vandijk11}.   Gusev \etal~\cite{gusev14,chuluunbaatar08,chuluunbaatar08a} have recently given an improved and extended application of the method discussed by Puzynin \etal~\cite{puzynin99,*puzynin00}.  They deal with the more general problem of a time-dependent Hamiltonian.  Using a truncated Magnus expansion with additional transformations, they are able to obtain stable and efficient solutions which are accurate up to sixth-order in the time step.      

Generally the various approaches involve time evolution and integration over space.  Thus there are a number of ways of dealing with the time evolution.    
Crank-Nicolson approximates the exponential time-evolution operator by a Cayley form which retains unitarity, but is correct only to low order in time advance~\cite{crank47}.  The Chebyshev polynomial expansion can lead to high accuracy even over significant time intervals.  It is not explicitly unitary.  The generalized Crank-Nicolson approximates the evolution operator with a $[M/M]$ Pad\'e approximant, factorized into $M$ factors of Cayley form.   This form is unitary and has a truncation error of $O[(\Delta t)^{2M+1}]$, where $\Delta t$ is the temporal step size.  This improves the precision rapidly with increasing $M$.  Besides these three approaches there are other approximations of the time-evolution operator, e.g., the exponential split-operator method~\cite{feit82} or the iterative Lanczos reduction~\cite{park86}.  Like time development the spatial integration can be achieved in different ways, e.g., by different types of finite differencing or by the pseudospectral fast Fourier transform approach. 

Since many of the calculations referred to have been done in one spatial dimension, in this paper we consider the generalized Crank-Nicolson with two spatial dimensions.  A number of articles have appeared recently that describe methods of solving the two-dimensional time-dependent Schr\"odinger equation, including those with time-dependent potentials and nonlinear terms.  See, for example, Refs.~\cite{tian10,xu12,gao16,symes16,wang16,zhang16}.  A number of these use the Cayley form for the time evolution operator.   We wish to employ the higher order Pad\'e form in order to enhance the efficiency of  the approach.  Given the two spatial dimensions, we pursue an alternating-direction implicit scheme which requires only solving one-dimensional implicit problems for each time step.  Different approaches have been suggested, such as the use of multigrid partitioning~\cite{gaspar14}, but it is our intention to present one that provides the user with another efficient alternative.  Clearly the method chosen will depend on the context.

In section~\ref{sec:2} we formulate the time dependence of the problem. Section~\ref{sec:3} is a description of the spatial integration.  A number of applications are discussed in Sec.~\ref{sec:4}, and Sec.~\ref{sec:5} presents conclusions and a discussion of the work. 

\section{Accurate time-evolution scheme}\label{sec:2}

We solve the two-dimensional time-dependent Schr\"odinger equation
\begin{equation}\label{eq:2.01}
\left(\hH-i\hbar\dfrac{\partial~}{\partial t}\right)\Psi(x,y,t) =0,
\end{equation}
where 
\begin{equation}\label{eq:2.02}
\begin{split}
\hH & = -\dfrac{\hbar^2}{2m}\dfrac{\partial^2~}{\partial x^2}  -\dfrac{\hbar^2}{2m}\dfrac{\partial^2~}{\partial y^2} + \hV(x,y) \\
& = \hK_x +\hK_y + \hV(x,y),
\end{split}
\end{equation}
starting with an initial wave function
\begin{equation}\label{eq:2.03}
\Psi(x,y,0) = \Phi(x,y).
\end{equation}

The time-evolution operator of the system  gives an expression for the wave function at a time in terms of the wave function at an earlier time,  i.e.,
\begin{equation}\label{eq:2.04}
\Psi(t+\Delta t) = e^{\ts -i\hH\Delta t/\hbar}\Psi(t),
\end{equation} 
where $\Delta t$ is the time advance, and where we have suppressed the spatial coordinates $x$ and $y$ in the wave function.
We will employ the factorized $[M/M]$ Pad\'e approximant along with  the alternating-direction implicit method~\cite{peaceman55}.  In keeping with the expansion of the time-evolution operator discussed in Ref.~\cite{vandijk07}, the operator is written as
\begin{equation}\label{eq:2.05}
e^{\ts -i\hH\Delta t/\hbar} = \prod_{s=1}^M \hTheta_s^{(M)} +  O[(\Delta t)^{2M+1}],
\end{equation}
where 
\begin{equation}\label{eq:2.06}
\hTheta_s^{(M)} \equiv \dfrac{1+(i\hH\Delta t/\hbar)/z_s^{(M)}}
{1-(i\hH\Delta t/\hbar)/\bar{z}_s^{(M)}},
\end{equation}
and $z_s^{(M)}, s=1,\dots,M$ are the roots of the numerator of the $[M/M]$ Pad\'e approximant of $e^{\ts z}$; the $\bar{z}_s^{(M)}$ are the corresponding complex conjugates.  Since $\Psi^{(n+1)} = e^{-i\hH\Delta t/\hbar}\Psi^{(n)}$ ($n$ refers to the time $t_n = n\Delta t, n=0,1,\dots$), we write
\begin{equation}\label{eq:2.07}
\Psi^{(n+1)} = \prod_{s=1}^M \hTheta_s^{(M)}\Psi^{(n)}.
\end{equation}
Defining $\Psi^{(n+s/M)}\equiv \hTheta_s^{(M)}\Psi^{(n+(s-1)/M)}$, we can solve for $\Psi^{(n+1)}$ iteratively starting with $\Psi^{(n+1/M)} = \hTheta_1^{(M)}\Psi^{(n)}$, then $\Psi^{(n+2/M)} = \hTheta_2^{(M)}\Psi^{(n+1/M)}$, and so on.  

Let us start with the basic substep of the procedure in going from $\Psi^{(n+(s-1)/M)}$ to $\Psi^{(n+s/M)}$, which we label below generically as $\Psi^0$ and $\Psi^+$, respectively.  We then write  
\begin{equation}\label{eq:2.08}
\Psi^+ = \left(\dfrac{1+(i\hH\Delta t/\hbar)/z}{1-(i\hH\Delta t/\hbar)/\bar{z}}\right)\Psi^0,
\end{equation}
or
\begin{equation}\label{eq:2.09}
\left(1-(i\hH\Delta t/\hbar)/\bar{z}\right)\Psi^+ = 
\left(1+(i\hH\Delta t/\hbar)/z\right) \Psi^0,
\end{equation}
where $z$ is the generic $z^{(M)}_s$.  Since $\hH = \hK_x+\hK_y + \hV$,  we write
\begin{equation}\label{eq:2.10}
\begin{split}
&\left[1-i(\hK_x+\hK_y+\hV)\dfrac{\Delta t}{\hbar\bar{z}}\right]\Psi^+ \\
& ~~~~ =
\left[1+i(\hK_x+\hK_y+\hV)\dfrac{\Delta t}{\hbar z}\right]\Psi^0 
\end{split}
\end{equation}
so that
\begin{widetext}
\begin{equation}\label{eq:2.11}
\begin{split}
\left(1-i\hK_x\dfrac{\Delta t}{\hbar\bar{z}}\right) & \left(1-i\hK_y\dfrac{\Delta t}{\hbar\bar{z}}\right)\Psi^+ +\hK_x\hK_y\dfrac{(\Delta t)^2}{\hbar^2\bar{z}^2}\Psi^+  \\ 
& =
\left(1+i\hK_x\dfrac{\Delta t}{\hbar\bar{z}}\right) \left(1+i\hK_y\dfrac{\Delta t}{\hbar\bar{z}}\right)\Psi^0 +\hK_x\hK_y\dfrac{(\Delta t)^2}{\hbar^2\bar{z}^2}\Psi^0 
 +i\hV\dfrac{\Delta t}{\hbar\bar{z}}\Psi^+ + i\hV\dfrac{\Delta t}{\hbar z}\Psi^0.
\end{split}
\end{equation}
In keeping with Peaceman and Rachford~\cite{peaceman55}, we define $\widetilde{\Psi}$ by the equation
\begin{equation}\label{eq:2.12}
\left(1-i\hK_x\dfrac{\Delta t}{\hbar\bar{z}}\right)\widetilde\Psi 
=\left(1+i\hK_y\dfrac{\Delta t}{\hbar{z}}\right)\Psi^0 + i\left(1-i\hK_x\dfrac{\Delta t}{\hbar\bar{z}}\right)\hV\dfrac{\Delta t}{\hbar z}\Psi^0.
\end{equation}
We insert this expression into Eq.~(\ref{eq:2.11}) to obtain
\begin{equation}\label{eq:2.13}
\begin{split}
\left(1-i\hK_x\dfrac{\Delta t}{\hbar\bar{z}}\right)\left(1-i\hK_y\dfrac{\Delta t}{\hbar\bar{z}}\right)\Psi^+ = & \left(1+i\hK_x\dfrac{\Delta t}{\hbar{z}}\right) \left(1-i\hK_x\dfrac{\Delta t}{\hbar\bar{z}}\right)\widetilde\Psi 
 -\left(1+i\hK_x\dfrac{\Delta t}{\hbar{z}}\right)\left(1-i\hK_x\dfrac{\Delta t}{\hbar\bar{z}}\right)i\hV\dfrac{\Delta t}{\hbar z}\Psi^0 \\
& -\hK_x\hK_y\dfrac{(\Delta t)^2}{\hbar^2\bar{z}^2}\Psi^+ + \hK_x\hK_y \dfrac{(\Delta t)^2}{\hbar^2 z^2}\Psi^0 
+ iV\dfrac{\Delta t}{\hbar\bar{z}}\Psi^+ + i\hV\dfrac{\Delta t}{\hbar z}\Psi^0.
\end{split}
\end{equation}
Operating on Eq.~(\ref{eq:2.13}) with the inverse of $\left(1-i\hK_x\dfrac{\Delta t}{\hbar\bar{z}}\right)$, we get
\begin{equation}\label{eq:2.14}
\begin{split}
\left(1-i\hK_y\dfrac{\Delta t}{\hbar\bar{z}}\right)\Psi^+ & = \left(1+i\hK_x\dfrac{\Delta t}{\hbar{z}}\right)\widetilde\Psi
 -\left(1+i\hK_x\dfrac{\Delta t}{\hbar{z}}\right) i\hV\dfrac{\Delta t}{\hbar z}\Psi^0 \\
& -\left(1-i\hK_x\dfrac{\Delta t}{\hbar\bar{z}}\right)^{-1}\hK_x\hK_y\dfrac{(\Delta t)^2}{\hbar^2\bar{z}^2}\Psi^+ +\left(1-i\hK_x\dfrac{\Delta t}{\hbar\bar{z}}\right)^{-1} \hK_x\hK_y \dfrac{(\Delta t)^2}{\hbar^2 z^2}\Psi^0 \\
&+\left(1-i\hK_x\dfrac{\Delta t}{\hbar\bar{z}}\right)^{-1} i\hV\dfrac{\Delta t}{\hbar\bar{z}}\Psi^+ +\left(1-i\hK_x\dfrac{\Delta t}{\hbar\bar{z}}\right)^{-1} i\hV\dfrac{\Delta t}{\hbar z}\Psi^0.
\end{split}
\end{equation}
The inverse operators are expanded, but we must make sure that the expansions are correct to $O[(\Delta t)^{2M}]$ since the overall expansion~(\ref{eq:2.05}) is of that order.

Simplifying and keeping terms up to $(\Delta t)^{2M}$ and assuming $M>1$, we obtain
\begin{equation}\label{eq:2.15}
\begin{split}
\left(1-iK_y\dfrac{\Delta t}{\hbar\bar{z}} - iV\dfrac{\Delta t}{\hbar\bar{z}}\right)\Psi^+ & = \left(1+iK_x\dfrac{\Delta t}{\hbar{z}}\right)\widetilde\Psi
 +\dfrac{(\Delta t)^2}{\hbar^2{z}^2}K_xV\Psi^0 \\
& -\sum_{l=0}^{2(M-1)} \left(iK_x\dfrac{\Delta t}{\hbar\bar{z}}\right)^l K_xK_y\dfrac{(\Delta t)^2}{\hbar^2\bar{z}^2}\Psi^+ +\sum_{l=0}^{2(M-1)} \left(iK_x\dfrac{\Delta t}{\hbar\bar{z}}\right)^l K_xK_y \dfrac{(\Delta t)^2}{\hbar^2 z^2}\Psi^0 \\
& +\sum_{l=1}^{2M-1} \left(iK_x\dfrac{\Delta t}{\hbar\bar{z}}\right)^l iV\dfrac{\Delta t}{\hbar\bar{z}}\Psi^+ + \sum_{l=1}^{2M-1} \left(iK_x\dfrac{\Delta t}{\hbar\bar{z}}\right)^l iV\dfrac{\Delta t}{\hbar z}\Psi^0 + O\left[(\Delta t)^{2M+1}\right].
\end{split}
\end{equation}
The $M=1$ case, for which $z=-2$, results in the equation
\begin{equation}\label{eq:2.16}
\left(1 +i\hK_y\dfrac{\Delta t}{2} + i\hV\dfrac{\Delta t}{2}\right)\Psi^+ = \left(1-i\hK_x\dfrac{\Delta t}{2}\right)\widetilde\Psi.
\end{equation}
This equation is a typical implicit equation with the Cayley form.
We solve Eq.~(\ref{eq:2.15}) iteratively so that $\Psi^{\sigma+1}\rightarrow\Psi^+$ as $\sigma=0,1,2,\dots$ increases in the equation 
\begin{equation}\label{eq:2.17}
\begin{split}
\left(1-i\hK_y\dfrac{\Delta t}{\hbar\bar{z}} - i\hV\dfrac{\Delta t}{\hbar\bar{z}}\right)\Psi^{\sigma+1} & = \left(1+i\hK_x\dfrac{\Delta t}{\hbar{z}}\right)\widetilde\Psi
 +\dfrac{(\Delta t)^2}{\hbar^2{z}^2}\hK_x\hV\Psi^0 \\
& -\sum_{l=0}^{2(M-1)} \left(i\hK_x\dfrac{\Delta t}{\hbar\bar{z}}\right)^l \hK_x\hK_y\dfrac{(\Delta t)^2}{\hbar^2\bar{z}^2}\Psi^\sigma +\sum_{l=0}^{2(M-1)} \left(i\hK_x\dfrac{\Delta t}{\hbar\bar{z}}\right)^l \hK_x\hK_y \dfrac{(\Delta t)^2}{\hbar^2 z^2}\Psi^0 \\
& +\sum_{l=1}^{2M-1} \left(i\hK_x\dfrac{\Delta t}{\hbar\bar{z}}\right)^l i\hV\dfrac{\Delta t}{\hbar\bar{z}}\Psi^\sigma + \sum_{l=1}^{2M-1} \left(i\hK_x\dfrac{\Delta t}{\hbar\bar{z}}\right)^l i\hV\dfrac{\Delta t}{\hbar z}\Psi^0.
\end{split}
\end{equation}
\end{widetext}
We start the iteration with setting $\Psi^{\sigma=0} =\Psi^0$.  When $\Psi^{\sigma +1}$ and $\Psi^\sigma$ are sufficiently close we stop.  Note that we need to calculate $\widetilde\Psi$ only once for each sequence of iterations.  We find that this approach can give accurate results; typically around six iterations are required for precise results.  This process has to be repeated for each of the $M$ steps needed to achieve a full time step advance.

There is an alternative approach to solving Eq.~(\ref{eq:2.15}) for $\Psi^+$.  The terms on the right side involving $\Psi^+$ can be moved to the left side and one solves a linear system of equations upon the discretization of the spatial variables.  However, as we show in the next section, the kinetic energy operators are banded diagonal matrices, and those operators raised to some power would result in matrices with the size of the bands increased.  As a result the gains in efficiency of a banded matrix formulation are lost.
   
\section{Spatial integration}\label{sec:3}
\setcounter{equation}{0}
The numerical spatial integration of the partial differential equation~(\ref{eq:2.17}) can be done in a number of ways.  Two approaches often considered are the spectral decomposition of the spatial (kinetic energy) operator or the finite-difference representation of this operator.  The relative merits are discussed by the authors of Ref.~\cite{cerjan91}.  They point out that a``low-order differencing method is in principle faster than a spectral method since it scales as the 'bandedness' times the size of the grid, $O(bN)$, rather than as $O(N\log N)$".  In the case of two-dimensional systems using the alternating-direction implicit approach $N$ is replaced by $N^2$, whereas $b$ is unchanged.      For the purpose of this work we therefore use finite differences.   One could choose the traditional three-point expression for the second-order partial derivative.  There are however more precise methods.  For instance the recent Numerov recent approach~\cite{vandijk16} gives much higher accuracy, as does the high-order compact finite difference approach in Refs.~\cite{tian10,xu12}.  The traditional approach is $O(h^2)$, where $h$ is the spatial step size, whereas the high-order compact method is $O(h^4)$, and the Numerov algorithm is $O(h^5)$.  The advantage of these approaches is that they lead to three-point formulas which may be convenient when crossing a discontinuity of the potential or considering an adaptive spatial grid~\cite{vandijk16}.  

As in earlier work~\cite{vandijk07} we consider formulas which allows one to choose an arbitrary order of $h$.  For a spatial grid (in one dimension) with step size $h$, the second derivative of $f(x)$ is expanded as
\begin{equation}\label{eq:3.01}
f''(x) =   \dfrac{1}{h^2}\sum_{k=-r}^{k=r} c_k^{(r)}f(x+kh) + O(h^{2r}),
\end{equation}
where the $c_k^{(r)}$ are real constants, obtained from making series expansions of the functions $f(x\pm kh)$.    A similar technique is used by Wang and Shao for the kinetic energy operator acting on the wave function of a two-dimensional stationary state problem~\cite{wang09}.  In another article the same authors suggest an expansion of the form~\cite{shao09}
\begin{equation}\label{eq:3.02}
\begin{split}
f''(x) = &\sum\limits_{\substack{k=-r \\ (k\ne 0)}}^{k=r}a_k^{(r)}f''(x+kh) \\
 &+\dfrac{1}{h^2}\sum_{k=-r}^{k=r} b_k^{(r)}f(x+kh) + O(h^{4r}).
 \end{split}
\end{equation}
In one dimension the discretized kinetic energy is expressed as a banded diagonal matrix with bandwidth of $2r+1$, just like in the case of Eq.~(\ref{eq:3.01}).  Thus it seems that with virtually the same effort the calculation gives much more accurate results.  A comparison of the two expansions~\cite{vandijk11} shows that for smaller values of $r$ the calculation is indeed much more efficient, however for larger $r$ the accuracy decreases.   The kinetic energy operator resulting from Eq.~(\ref{eq:3.01}) can be made strictly diagonally dominant, whereas the diagonal dominance of the kinetic energy matrix from Eq.~(\ref{eq:3.02}) becomes compromised when $r$ goes beyond ten.  In this paper we use expansion~(\ref{eq:3.01}) for the kinetic energy operators $\hK_x$ and $\hK_y$. 

We consider a rectangular domain in space $[x_0,x_\cJ]\times [y_0,y_\cM] \subset \mathbb{R}^2$, which we partition uniformly in each direction, so that with $h_x = (x_\cJ-x_0)/\cJ$ and $h_y = (y_\cM-y_0)/\cM$, $x_j=x_0+jh_x, j=0,\dots,\cJ$ and $y_m=y_0+mh_y, m=0,\dots,\cM$.  The time is also partitioned over the time interval from 0 to $T$ into $N$ subintervals, so that  $\Delta t = T/N$ and the intermediate times are $t_n=n\Delta t$, where $n=0,1,\dots,N$.  The equations we need to solve are typically of the type Eqs.~(\ref{eq:2.12}) and (\ref{eq:2.17}).  If we let $\Psi(x_j,y_m)=\Psi_{j,m}$, then $(\hV\Psi)_{j,m}=V_{j,m}\Psi_{j,m}$ and 
\begin{equation}\label{eq:3.03}
\begin{split}
(\hK_x\Psi)_{j,m} = & \left(-\dfrac{\hbar^2}{2m}\right)\dfrac{1}{h_x^2}\sum_{k=-r}^{k=r}c_k^{(r)}\Psi_{j+k,m} \\
& \hspace{1in} \ \ \mathrm{for} \ \ \ 0\le j+k \le \cJ.
\end{split}
\end{equation} 
There is a similar relation for $\hK_y\Psi$ except that the summation is over the second index of $\Psi_{j,m+k}$.  Thus in Eq.~(\ref{eq:2.17}), for example, the right side is completely specified, but the $\Psi^{\sigma+1}$ on the left side needs to be found.  This equation is really a linear system of equations with a banded diagonal coefficient matrix over the index $m$.  It can be solved for each $j$ to obtain $\Psi^{\sigma+1}$.  It is the strength of the alternating-direction implicit scheme that calculations are reduced to one-dimensional ones.

\section{Implementation}\label{sec:4}

In this section we consider four examples in which the method outlined previously is applied in order to investigate its accuracy and efficiency.  We will also demonstrate the feasibility of calculating wave functions with more complex structure (several peaks and valleys) as they evolve in time.
    
\setcounter{equation}{0}
\subsection{Errors}

The truncation error of the series expansion of the wave function in time and space can be expressed as
\begin{equation}\label{eq:4.00a}
e = e^{(M)} + e^{(r)} = C_1 (\Delta t)^{2M+1} + C_2 h^{2r},
\end{equation}
where we are considering the error of the real quantity $|\Psi(x,y,t)|$ and $C_1$ and $C_2$ are real positive numbers related to the $(2M+1)$th partial derivative with respect to $t$ and the $(2r)$th derivative with respect to $x$ or $y$, respectively.  For simplicity we assume $h_x=h_y=h$ and a $xy$ symmetry of the wave function.  If exact analytic solutions are available the error can be calculated by comparison.  If that is not the case, a good estimate of the error can be made by comparing the solution for particular $M$ and $r$ to the one obtained when one or both of the $M$ and $r$ are increased by unity~\cite{vandijk11}.

To make comparisons of the  numerically obtained solutions to analytic solutions in cases where the latter are known,  we define the error $e_2$ such that
\begin{equation}\label{eq:4.01}
e_2^2 = \int_{x_0}^{x_\cJ}\int_{y_0}^{y_\cM} |\Psi(x,y,T)-\Psi_\mathrm{exact}(x,y,T)|^2\; dydx.
\end{equation}
The error $e_2$, which is a Euclidean/$\ell_2$-vector norm, is a measure of the accuracy of the wave function and its phase.  Alternatively some authors have used the $\ell_\infty$-vector norm
\begin{equation}\label{eq:4.01a}
e_\infty = \max_{j,m} |\Psi_{j,m} - \Psi_{\rm exact}(x_j,y_m,t)| \ \ \mathrm{for} \ \ t=T.
\end{equation}
In the case that no exact solution is available, one can make an estimate of the error by comparing a solution obtained with particular values of $M$ and $r$ to the solution obtained with $M+1$ and $r+1$, e.g.,
\begin{equation}\label{eq:4.01bb}
\eta_2^2 = \int_{x_0}^{x_\cJ}\hspace{-0.1in}\int_{y_0}^{y_\cM} |\Psi^{(M,r)}(x,y,T)-\Psi^{(M+1,r+1)}(x,y,T)|^2\; dydx.
\end{equation}
Since it turns out that $e_2$ and $\eta_2$ are very similar, evaluating $\eta_2$ provides a method to estimate the accuracy in the absence of an analytic solution~\cite{vandijk11,vandijk16}.

Since in our applications the wave functions are zero near the boundary of the domain, we can use the simple rectangle rule for integration.  The corrections to higher order polynomial approximations are all in terms of evaluations of the integrand near the end points, but since the wave function is zero there, one gets very accurate integrals with the simple quadrature~\cite{peters68}.

\subsection{Example 1: solvable two-dimensional potential}

\begin{figure}[h]
\resizebox{3.5in}{!}{\includegraphics[angle=-0]
{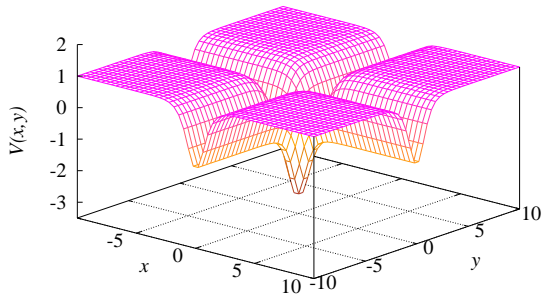}} 
\caption{(Colour online)  The potential function of example 1, Eq.~(\ref{eq:4.01b}).}
\label{fig:00a}
\end{figure}

Consider the potential 
\begin{equation}\label{eq:4.01b}
V(x,y) = -\dfrac{\hbar^2}{2m}(3-2\tanh^2 x - 2 \tanh^2 y)
\end{equation}
with $\hbar=2m=1$, see Fig.~\ref{fig:00a}.
This potential has been used by several authors as one which tests numerical methods~\cite{ciegis13,tian10,xu12}.  We choose the domain $[-20,20]\times [-20,20]$ and solve the problem from $t=0$ to $t=T=1$ with the solution on the boundary equal to zero~\cite{ciegis13}.  The exact solution is
\begin{equation}\label{eq:4.01c}
\Psi(x,y,t) = \dfrac{i e^{\ts it}}{2\cosh{x}\cosh{y}},
\end{equation}
which is also used to determine the initial wave function.  It should be noted that wave function (\ref{eq:4.01c}) is square integrable and is an energy eigenstate with energy $-\hbar^2/(2m)$.  It describes a bound state at threshold; the energy spectrum at higher energies is a continuum and the corresponding wave functions are unbound.

\begin{figure}[h]
\resizebox{3.5in}{!}{\includegraphics[angle=-0]
{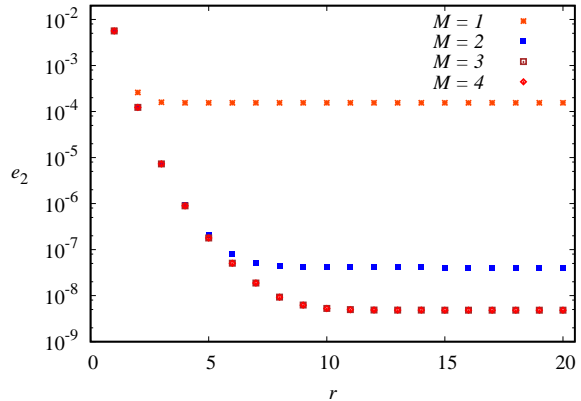}} 
\caption{(Colour online)  The errors $e_2$ of the numerical wave functions for potential (\ref{eq:4.01}).  The parameters are $T=1$ and $N=100$.}
\label{fig:00}
\end{figure}

In our numerical calculation we allow $M=1,\dots,6$ and $r=1,\dots,20$, with
$\Delta t = 0.01$ (or 100 time steps) and with ${\cal J}={\cal M} = 200$.  The $e_2$ are plotted as a function of $r$ for various values of $M$ in Fig.~\ref{fig:00}. 

We compare this calculation to that of Ref.~\cite{ciegis13} since the other two calculations~\cite{tian10,xu12} are done over a much smaller spatial domain with much smaller time intervals.  In our calculation we find generally that $e_2\gtrsim e_\infty$.  In Ref.~\cite{ciegis13} the quoted errors are approximately $e_\infty\approx 10^{-4}$.   Figure~\ref{fig:00} shows that the error in the calculation is reduced significantly when one goes from $M=1$ to $M=2$ and 3.  When $M>3$ the results are identical to those of $M=3$.  Given that earlier calculations referred to are basically $M=1$ calculations with spatial errors of the order of $h^3$ or $h^5$, this method results in significant improvement in accuracy. 

The errors of this example for $M\geq 3$ saturate at $e_2\approx 1.5\times 10^{-9}$.  In the ``gullies" of potential (\ref{eq:4.01b}) the magnitude of the wave function is larger than elsewhere.  In the gullies at the boundary of the computational space it is approximately $10^{-9}$.  The numerical calculation assumes that the wave function is zero outside the computational domain.  The discrepancy between the numerical wave function and the exact one outside the computational space is the source of the residual error. 

In order to determine the efficiency of the calculation we obtain the CPU time when the error is close to, but less than, $10^{-6}$.  For a particular $M$ we adjust the spatial step size by choosing $r$ until we reach a minimum CPU time.  Similarly for a particular $r$, we chose $M$ to yield minimum CPU time.   The results are listed in Tables~\ref{table:cpu_r} and \ref{table:cpu_M}.
In Ref.~\cite{vandijk07} we give an estimate for the CPU time as a function of $r$ for the one-dimensional calculation.  In two-dimensions we expect the behaviour to be similar since errors in $x$ and $y$ integrations are similar and additive, especially in symmetric cases.  Thus,
\begin{equation}\label{eq:4.01d}
\mathrm{CPU~time} \propto \left(e^{(r)}\right)^{-1/(2r)}r.
\end{equation}
To obtain a formula for the CPU time as a function of $M$ when $r$ and the error are constant, we take the relation of the error and $M$ to be
\begin{equation}\label{eq:4.01e}
e^{(M)} \propto (\Delta t/M^\nu)^{2M+1},
\end{equation}
where $\nu$ is a number less than unity.  In the one-dimensional case we showed that $\nu \approx 1$, but Table~\ref{table:cpu_M} shows that a better relationship, especially  for larger $M$, has $\nu<1$.  Since $T=N\Delta t$ is a constant we write
\begin{equation}\label{eq:4.01f}
\mathrm{CPU~time} \propto NM^2 \propto \left(e^{(M)}\right)^{-1/(2M+1)}M^{2+\nu}.
\end{equation}

\begin{table}[t]
\centering
\caption{Summary of CPU time as $r$ is varied assuming a constant error for example 1.  The fixed parameters are $M=5$, $N=100$, and $\Delta t=0.01$.}
\begin{tabular}{cccccccc}\hline\hline
$r$      &$~~~{\cal J}~~~$       &$~~~~h~~~~$         &CPU(s)   &$~e_2(10^{-7})~$\\ \hline
3      &290     &0.1379    &897    &8.27\\
4      &200     &0.1000   &396    &8.94\\
5      &165     &0.2424   &317    &9.97\\
6      &150     &0.2667   &302    &9.27\\
7      &140     &0.2587   &306    &9.63\\
8      &135     &0.2963   &317    &8.39\\
9      &130     &0.3077   &331    &9.96\\
10     &130    &0.3077   &364    &7.59\\
11     &125    &0.3200   &353    &8.27\\
12     &123    &0.3253   &392    &6.37\\
13     &123    &0.3253   &424    &4.98\\
14     &123    &0.3253   &457    &3.94\\
15     &122    &0.3279   &466    &9.75\\
16     &122    &0.3279   &535    &7.78\\
17     &122    &0.3279   &577    &7.14\\
18     &120    &0.3333   &581    &9.33\\
19     &120    &0.3333   &629    &8.62\\
20     &120    &0.3333   &693    &7.98\\  \hline\hline
\end{tabular}
\label{table:cpu_r}
\end{table}

\begin{figure}[h]
\resizebox{3.5in}{!}{\includegraphics[angle=-0]
{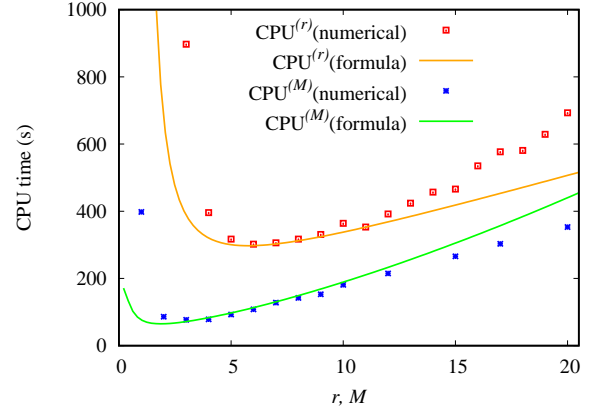}} 
\caption{(Colour online)  CPU time as function of $r$ and $M$ for calculation with potential~(\ref{eq:4.01b}).}
\label{fig:cpu}
\end{figure}

\begin{table}[t]
\centering
\caption{Summary of CPU time when $M$ is varied for example 1.  The fixed parameters are $r=6$, ${\cal J}=150$, and $h=0.2667$.}
\begin{tabular}{crccc}\hline\hline
$M$         &\multicolumn{1}{c}{$~~~N~~~$}       &$~~~~\Delta t~~~~$     &CPU(s)   &$e_2 (10^{-7})$   \\ \hline
1           &2500   &0.0004 &398    &10.00   \\
2          &130     &0.0077 &86.0     &9.36   \\
3           &30      &0.0333 &76.8   &9.74  \\
4           &25      &0.0400   &78.4   &9.35  \\
5           &20      &0.0500  &92.6   &9.27   \\
6           &18      &0.0556 &108    &9.28   \\
7          &15      &0.0667 &128    &9.28  \\
8          &14      &0.0714 &142    &9.28   \\
9          &12      &0.0833 &153    &9.27  \\
10        &11      &0.0909 &181    &9.27   \\
12        &10      &0.1000    &215    &9.27   \\
15        &8     &0.1250  &266    &9.27   \\
17         &7    &0.1429 &303    &9.27   \\
20         &6    &0.1667 &353    &9.27   \\ \hline\hline
\end{tabular}
\label{table:cpu_M}
\end{table}
The CPU times as a function of $r$ and $M$ are shown in Fig.~\ref{fig:cpu}.
The estimates of the errors are shown as solid lines.  For the CPU time as a function of $M$ we have estimated $\nu=1/2$.  Such an estimate seems reasonable in light of the fact that the iterative part of the procedure increases the time, and the number of iterations vary with the value of $M$.

\subsection{Example 2: oscillating and pulsation harmonic oscillator wave functions}

Consider the potential function for the two-dimensional anisotropic harmonic oscillator,
\begin{equation}\label{eq:4.02}
V(x,y) = \dfrac{1}{2}m(\omega_x^2x^2 + \omega_y^2y^2). 
\end{equation}
An analytic solution for such a potential is~\cite{vandijk14} 
\begin{equation}\label{eq:4.03}
\Psi(x,y,t)=\psi_{n_x}(\alpha_x,\beta_x;x,t)\psi_{n_y}(\alpha_y,\beta_y;y,t),
\end{equation}
where
\begin{equation}\label{eq:4.04}
\begin{split}
\psi_{n}(\alpha,\beta;x,t) = & \left(\dfrac{\alpha^2\beta}{\sqrt{\pi}2^{n}{n!}}\right)^{1/2}\dfrac{e^{\ts -i({n}+1/2)\theta}}{f^{1/4}} \\
& \times H_{n}(\xi)e^{\ts -\xi^2/2+i\cT}.
\end{split}
\end{equation}
The various quantities in Eq.~(\ref{eq:4.04}) are defined as follows:
\begin{equation}\label{eq:4.05}
\begin{split}
\alpha & =\sqrt{m\omega/\hbar}, \ \ \ f = \alpha^4\cos^2\omega t + \beta^4\sin^2\omega t \\
\xi & = \beta[\alpha^2(x-A\cos\omega t) - k\sin\omega t] /f^{1/2} \\
\cT & = \alpha^2/(2f)\left\{[(\beta^4 - \alpha^4)x^2 - k^2 + \beta^4A^2]\sin\omega t\cos\omega t \right. \\
 &~~~\left.+ 2[\alpha^4k x\cos\omega t + \beta^4A(k\sin\omega t -\alpha^2 x)\sin\omega t]\right\} \\
\theta & = \arctan\left(\dfrac{\beta^2\sin\omega t}{\alpha^2\cos\omega t}\right)+2\pi\nu, \ \ \nu = \mathrm{int}\left(\dfrac{\omega t + \pi}{2\pi}\right).
\end{split}
\end{equation} 
The wave function (\ref{eq:4.04}) is the pulsating and oscillating wave function of a particle subject to a one-dimensional harmonic oscillator  characterized by $\omega$ or $\alpha$.  The initial ($t=0$) wave function is the $n$th energy state of the particle subject to an oscillator characterized by $\beta$, rather than $\alpha$, displaced from the origin by amount $A$ and with a momentum $\hbar k$.  The function $H_{n}(\xi)$ is the $n$th-order Hermite polynomial.    This wave function provides a wave packet with more fluctuation than the traditional coherent wave packet; for instance, it has nodes which travel with the packet and whose occurrence spread and contract in time.

\begin{figure}[h]
\resizebox{3.5in}{!}{\includegraphics[angle=-0]
{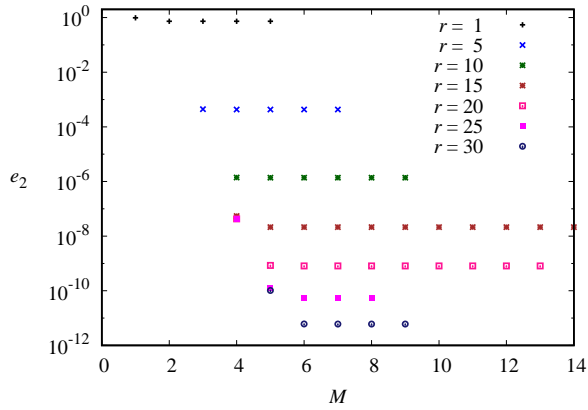}} 
\caption{(Colour online)  The errors of the numerical wave functions as they relate to the order of the Pad\'e approximant for various orders of the spatial expansion.  The parameters used are listed in Table~\ref{table:1}.}
\label{fig:01}
\end{figure}

As an initial study of the accuracy of the method we consider the simplest case of an isotropic oscillator with the initial state the ground state.  The values of the parameters are give in Table~\ref{table:1}.
\begin{table}[h]
\caption{Parameters used for the error calculations of Figs.~\ref{fig:01} and \ref{fig:02}.}
\begin{tabular}{l|l}
\hline\hline
$\hbar = m = 1, \ \ n_x=n_y=0$ & $k_x = k_y =0$ \\
$x_0 = y_0 = -15, \ \ x_\cJ=y_\cM=15$ & $\alpha_x=\beta_x=\alpha_y= \beta_y = 1$ \\
$t_\mathrm{max} = 2\pi, \ \ N = 100, \ \ dt = 2\pi/N$ & $A_x=A_y =2$  \\
$\cJ=\cM=100$ & \\
\hline\hline
\end{tabular}
\label{table:1} 
\end{table}
In Fig.~\ref{fig:01} the error as a function of $M$, the order of the diagonal Pad\'e approximant, is displayed.  Note that the lower values of $M$, especially for larger $r$ give no results because the convergence of the iterative part of the calculation is not achieved.  By decreasing $dt$ convergence can again be attained, but in Fig.~\ref{fig:01} we keep $dt$ constant .  The horizontal plateaux are not completed since there is no change in the error as $M$ is further increased.   The calculations are done with double-precision floating-point arithmetic.   We achieve an error less than $10^{-11}$ for $r=30$ and $M\ge 6$. The errors could be further reduced by increased computational precision. 

\begin{figure}[h]
\resizebox{3.5in}{!}{\includegraphics[angle=-0]
{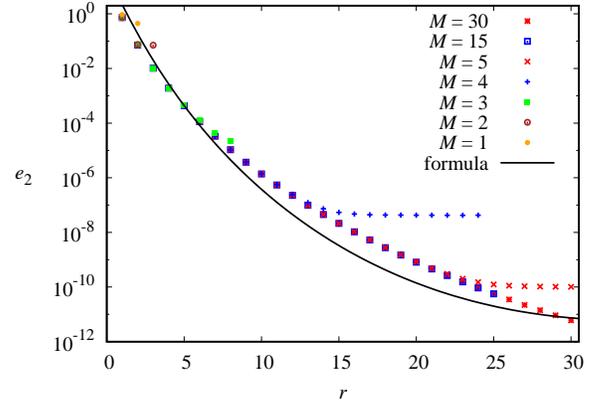}} 
\caption{(Colour online)  The errors of the numerical wave functions as a function of $r$ for different values of $M$.  The parameters used are listed in Table~\ref{table:1}.}
\label{fig:02}
\end{figure}

For the same model we plot the error as a function of $r$ in Fig.~\ref{fig:02} for a number of values of $M$.
The lower pattern of dots is one that is obtained for each $M$ value up to a particular value of $r$, at which the error is constant as $r$ is increased further.  These horizontal plateaux only extend to a certain point after which the iterative procedure becomes unstable.  The plateaux are clearly visible for $M=4$ and 5, and the beginnings can be discerned for $M=1,2,3$.  The instability of the calculation does not mean that we cannot obtain results in those regions.  In this calculation $dt$ is the same in all cases; where instability sets in a smaller $dt$ will restore stability.
The fact that the curves superimpose on the left can be seen from Fig.~\ref{fig:01} where for each $M\ge 6$ the errors converge for $r$ sufficiently large. 

On Fig.~\ref{fig:02} we have plotted a solid line which is an estimate of the error obtained by considering the truncation error of the expansion in $x$ or $y$, i.e.,
\begin{equation}\label{eq:4:05a}
e_2^{(r)} = C_2h^{2r} \approx \max_{y} \left|\dfrac{1}{(2r)!}\dfrac{\partial^{2r}~}{\partial x^{2r}}\Psi(x,y,T)|_{x=\xi}\right| h^{2r} \times \dfrac{1}{2.5^{2r}} ,
\end{equation}
where $\xi$ is some value in the domain of $x$.  We have  assumed that $x$ dependence of the wave function of this example is a Gaussian and that the maximum value of it and its even-order derivatives occur when the argument is zero~\cite[p. 933]{abramowitz65}, i.e.,
\begin{equation}\label{eq:4:05b}
G(x) = e^{-x^2/2} , \ \ \ G^{(2r)}(0) = \dfrac{(-1)^r(2r)!}{2^r(r!)}.
\end{equation}
The last factor on the right side of Eq.~(\ref{eq:4:05a}) is an adjustment to give reasonable agreement with the data.  It amounts to an effective spatial step size which is smaller by a  factor of 2.5.  The shape of the solid curve is very sensitive to the form of this factor.  

\begin{figure*}[t]
$\begin{array}{cc}
\resizebox{3.5in}{!}{\includegraphics[angle=-0]{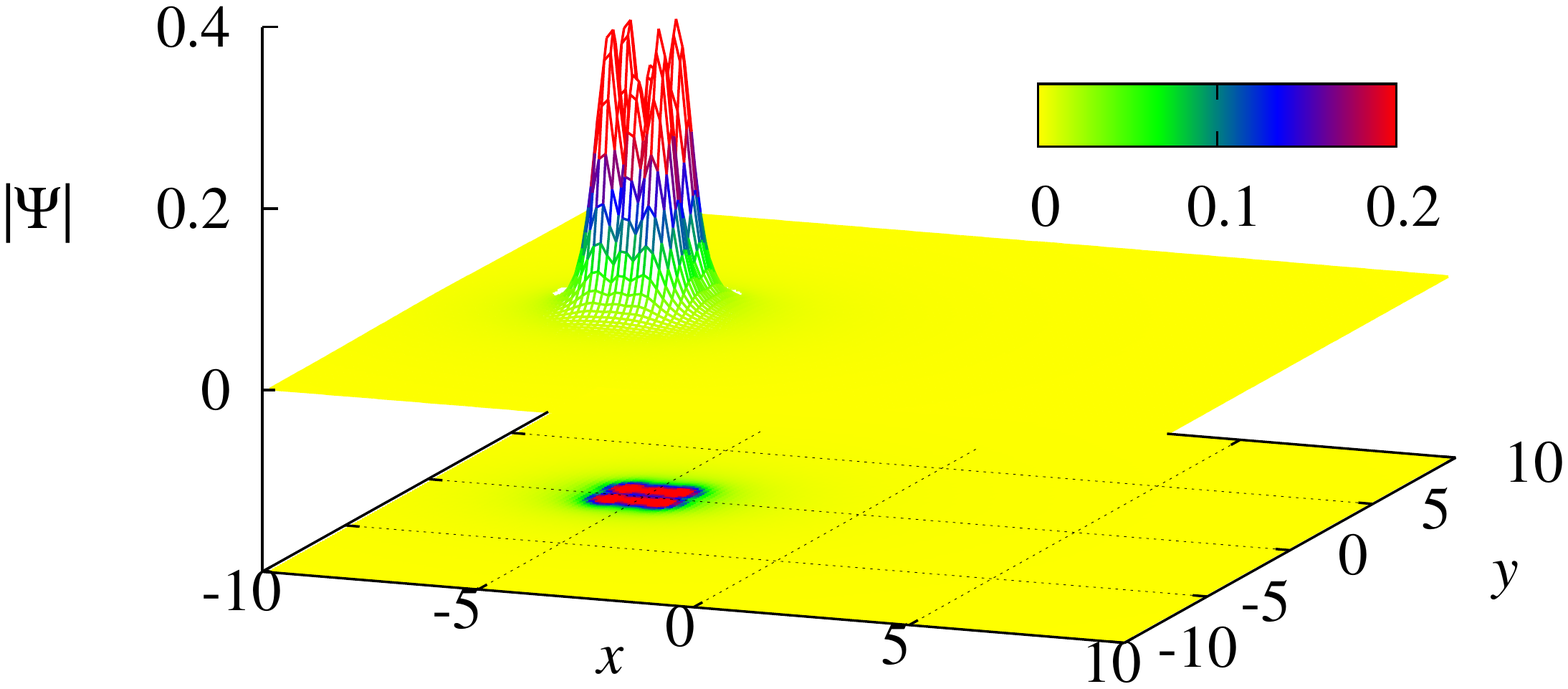}} 
& \resizebox{3.5in}{!}{\includegraphics[angle=-0]{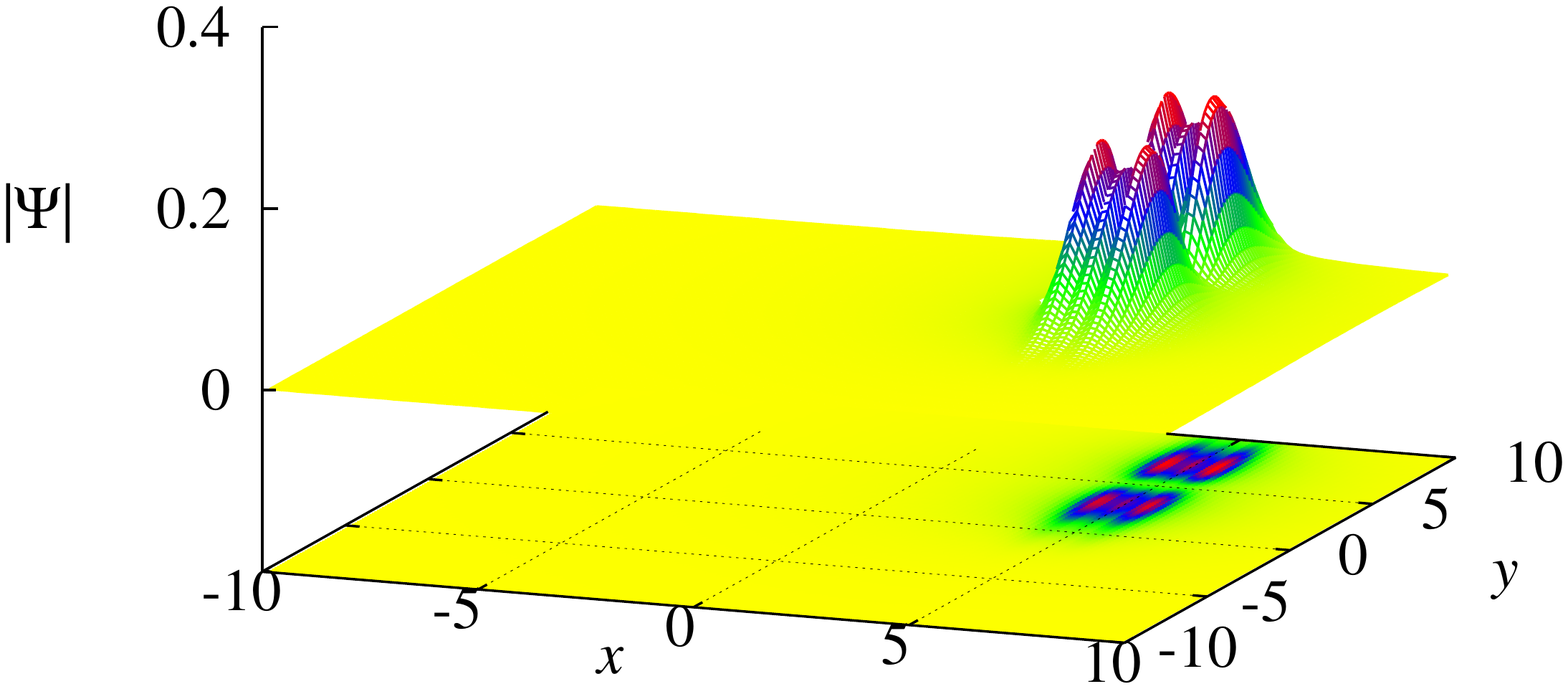}} \\
\resizebox{3.5in}{!}{\includegraphics[angle=-0]{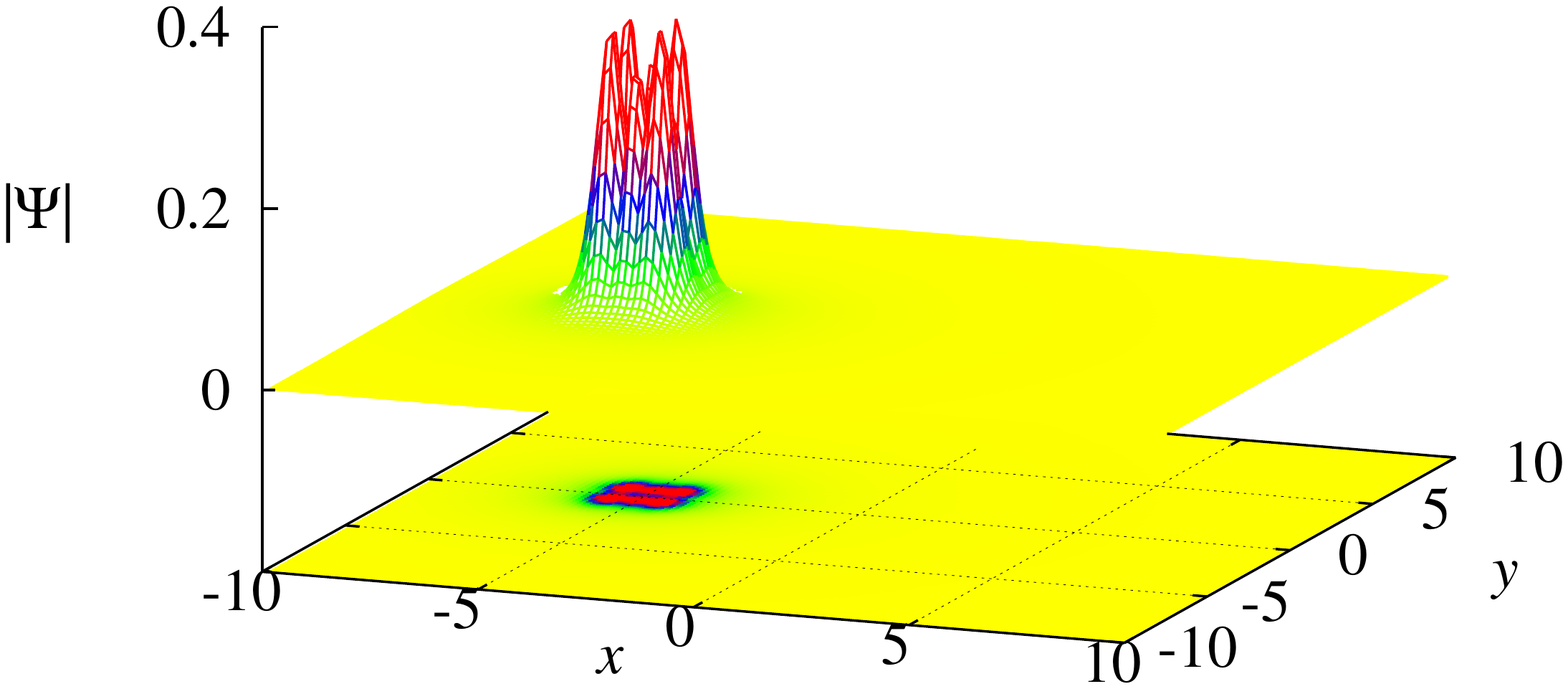}} 
&  \resizebox{3.5in}{!}{\includegraphics[angle=-0]{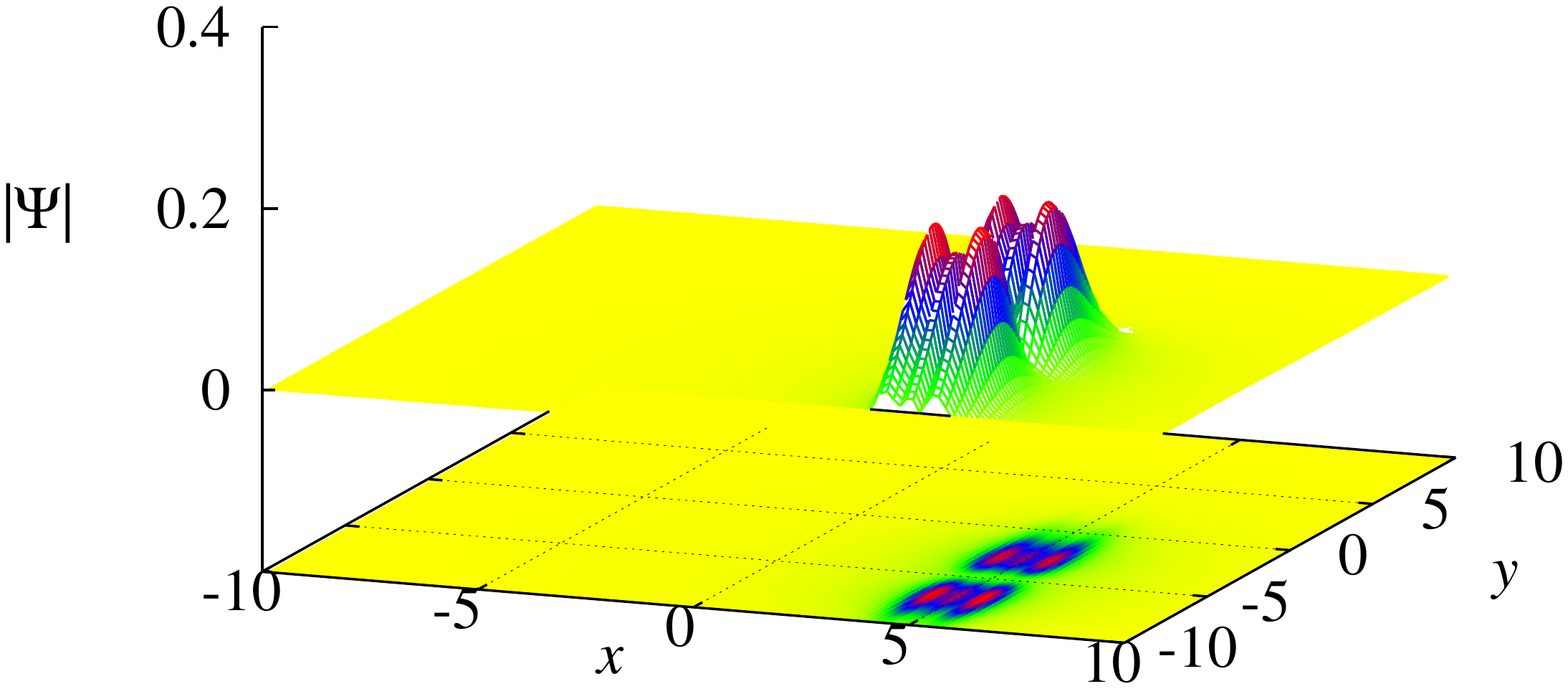}} \\
\end{array}$
\caption{(Colour online) Oscillating and pulsating wave packet in two dimensions of example 2.  The plot in the upper left panel represents the wave function at time $t=0$, in the upper right panel at $t=\pi/2$, in the lower left panel at $t=\pi$, and in the lower right panel at $t=3\pi/2$.   See animation~\href{http://physwww.physics.mcmaster.ca/~vandijk/fig:03.mp4}{here}.
}
\label{fig:03}
\end{figure*}

We plot the progression of the oscillating and pulsating wave packet as numerically determined in Fig.~\ref{fig:03}.  The parameters used for this calculation are listed in Table~\ref{table:2}.  The error $e_2$ in the calculation
\begin{table}[h]
\caption{Parameters used for calculation of the oscillating, pulsating wave function shown in Fig.~\ref{fig:03}.}
\begin{tabular}{l|l}
\hline\hline
$M=4, \ r=14,  \ n_x = 2, \  n_y = 1$ & $\hbar = m = 1, \ k_x = 0, \ k_y = 5$ \\
$x_0 = y_0 = -15, \ \ x_\cJ=y_\cM=15$ & $\alpha_x=\sqrt{2}, \ \ \ \alpha_y = 1$  \\
$t_\mathrm{max} = 2\pi, \ N = 557, \ dt = 2\pi/N$ & $\beta_x = 2\alpha_x, \ \ \ \beta_y=2\alpha_y$  \\
$\cJ=\cM=200$ & $A_x= -5, \ \ \ A_y =0$ \\
\hline\hline
\end{tabular}
\label{table:2} 
\end{table}
ranges from $3\times 10^{-7}$ after one time step to $5\times 10^{-3}$ after 557 steps.
Note that the oscillating frequency is $\omega_x = 2\omega_y$.  This means that when the motion has executed a complete cycle in the $x$ direction it has only gone through half a cycle in the $y$ direction.  Thus the packet starts at (-5,0), travels along a quarter-elliptical path to (5,5), then back to (-5,0), to (5,-5), and to the initial point (-5,0).  The pulsating frequencies in each direction are four times the corresponding oscillating frequencies.  

\subsection{Example 3: free wave packet}
For the free wave packet we consider the Hermite-Gaussian wave function of Ref.~\cite{vandijk14},
\begin{equation}\label{eq:4.06}
\Psi(x,y,t) =\psi(\alpha_x,n_x; x,t)\psi(\alpha_y,n_y;y,t) ,
\end{equation}
where 
\begin{equation}\label{eq:4.07}
\begin{split}
\psi(\alpha,n;z,t) =& \dfrac{N_n(\alpha)}{\sqrt{1+i\alpha^2\tau}}\exp\left(\dfrac{i(z-A)^2}{2\tau}\right)e^{\ts -in\theta} \\
& \times H_n(\xi)\exp\left(-\dfrac{\xi^2}{2} - i\dfrac{\xi^2}{2\alpha^2\tau}\right)
\end{split}
\end{equation}
with
\begin{equation}\label{eq:4.08}
\begin{split}
& \xi =\dfrac{\alpha[(z-A)-k\tau]}{\sqrt{1+\alpha^4\tau^2}}, \ \ \ \theta = \arctan(\alpha^2\tau),  \\
& N(\alpha)  = \left(\dfrac{\alpha}{\sqrt{\pi}2^nn!}\right)^{1/2} \ \ \ \mathrm{and} \ \ \ \tau = \hbar t/m.
\end{split}
\end{equation}
The travelling wave packet will have nodes whose distribution, if there is more than one node, spread in time.    The model is similar to that of Galbraith \etal~\cite{galbraith84} in whose calculation $n_x=n_y=0$ and $\alpha_x=\alpha_y$.   The parameters we use are given in Table~\ref{table:3}. 
\begin{table}[h]
\caption{Parameters used for the calculations of the free packet in Fig.~\ref{fig:04}.}
\begin{tabular}{l|l}
\hline\hline
$\hbar = 2m = 1$, \ \ $n_x=2, \ \ n_y=1$  & $k_x = k_y =32$ \\
$x_0 = y_0 = -2.5, \ \ x_\cJ=y_\cM=2.5$ & $\alpha_x=\alpha_y$ \\
$N = 80, \ \ dt = 10^{-4}$ & $A_x=A_y =0.25$  \\
$\cJ=\cM=200, \ \ r =20, \ \ M = 10$ & $\alpha_x = 12.5\sqrt{2}$ \\
\hline\hline
\end{tabular}
\label{table:3} 
\end{table}

\begin{figure*}[t]
$\begin{array}{cc}
\resizebox{3.5in}{!}{\includegraphics[angle=-0]{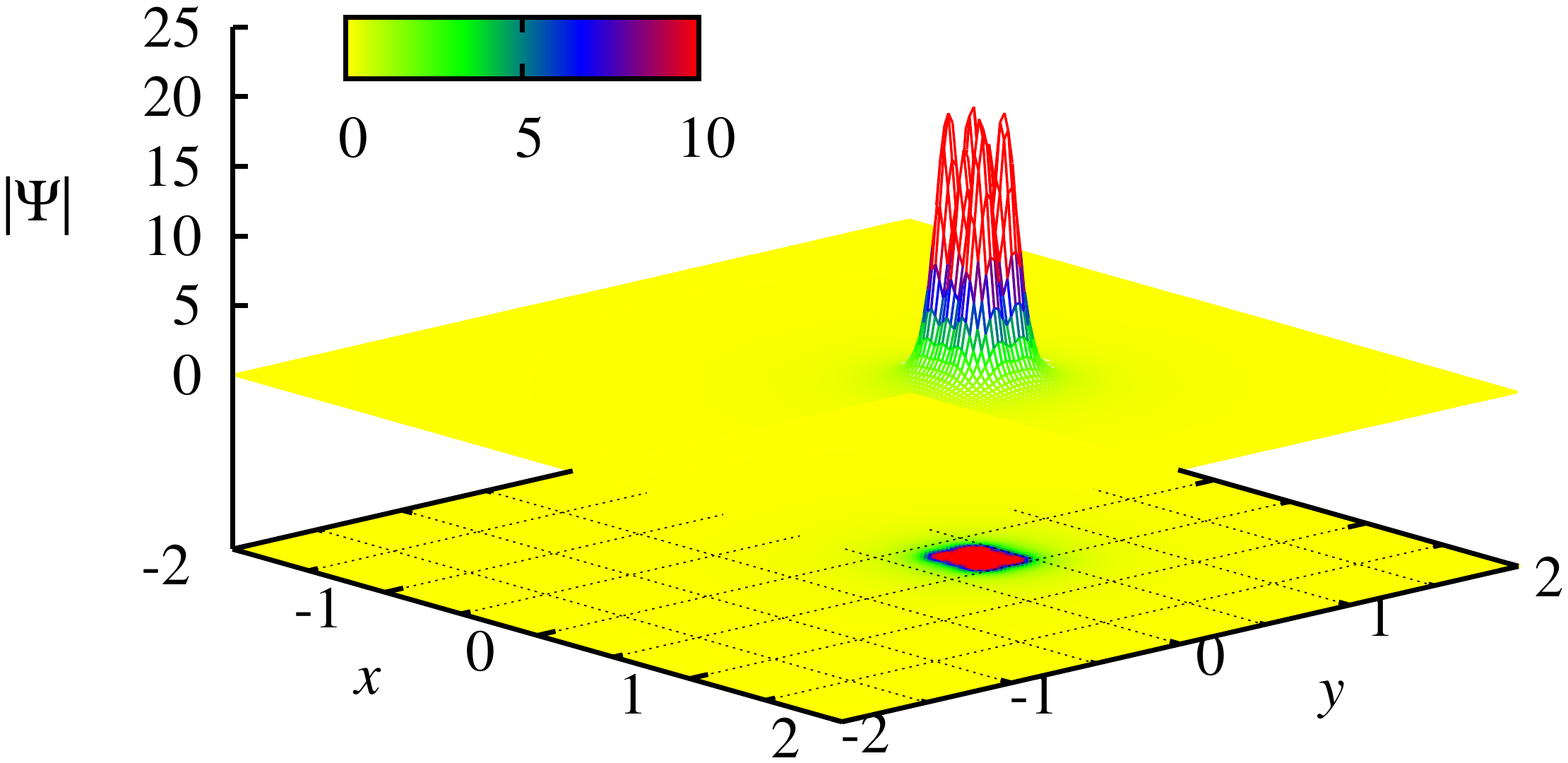}} &
\resizebox{3.5in}{!}{\includegraphics[angle=-0]{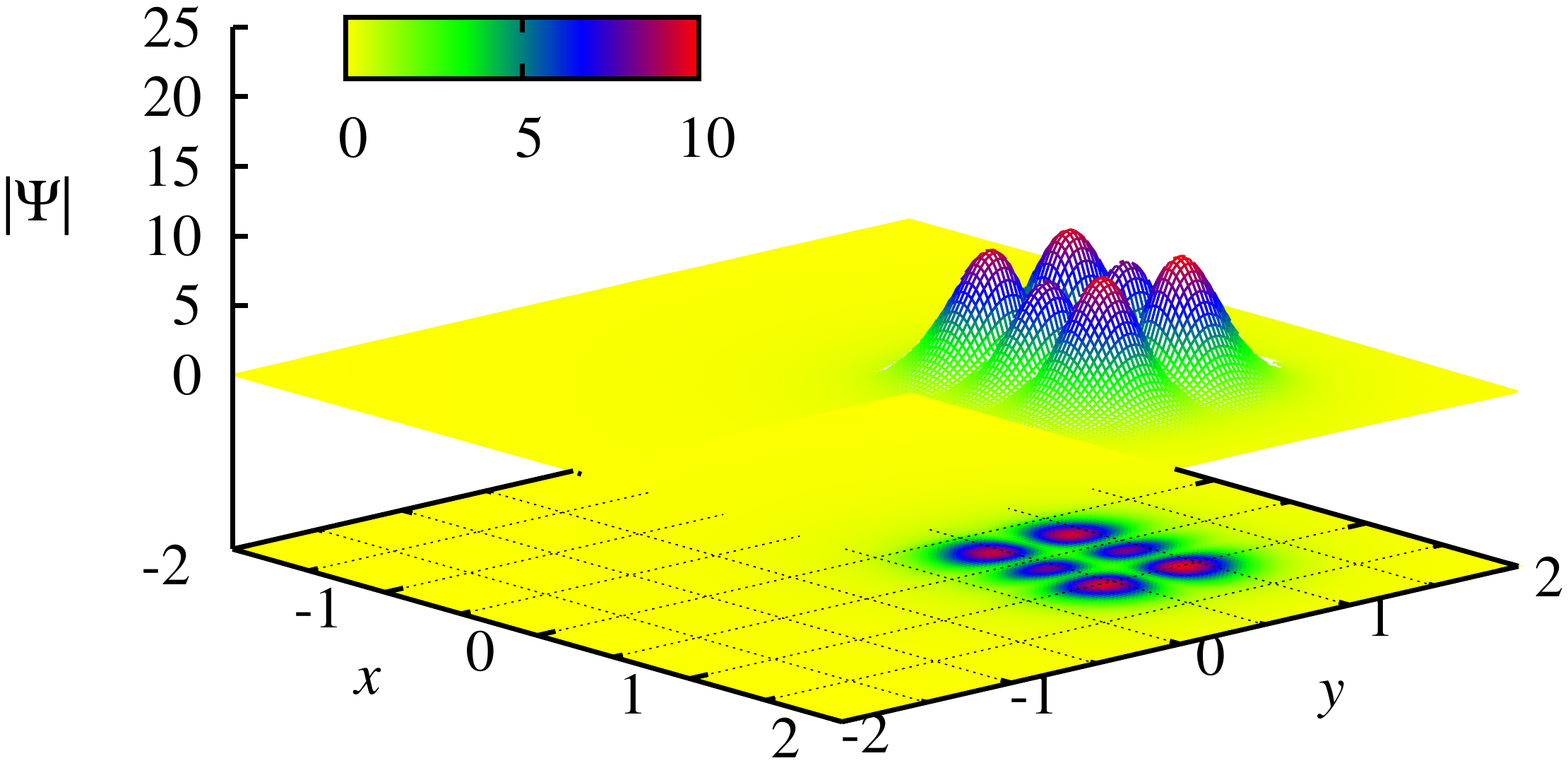}}
\end{array}$
\caption{(Colour online) The movement and dispersion of a free wave packet in example 3 shown at $t=10^{-3}$ in the left panel and at $t=6.3\times 10^{-3}$ in the right panel. See animation~\href{http://physwww.physics.mcmaster.ca/~vandijk/fig:04.mp4}{here}.
}
\label{fig:04}
\end{figure*}

The free wave packet at times $t=10^{-3}$ and $t=6.3\times 10^{-3}$ is shown in Fig.~\ref{fig:04}.  The separation of the peaks of the wave function as time progresses is clearly evident. 


\subsection{Example 4: single-slit diffraction}

The wave nature of electrons has been studied and observed in semiconductor nanostructures.    Endoh \etal~\cite{endoh92,*endoh99} have considered numerical simulations of the passage of such electrons through narrow constrictions.  Recent experiments observed controlled electron diffraction for both single- and double-slit configurations~\cite{bach13,khatua14}.

The single slit in the barrier is obtained by introducing a potential
\begin{equation}\label{eq:4.09}
V(x,y) = V_0 f(x) [f(0)-f(y)],
\end{equation}
where $f(x)$ is the difference of two Fermi functions
\begin{equation}\label{eq:4.10}
f(x) = \dfrac{1}{1+e^{\ts -\mu(x+x_w)}} - \dfrac{1}{1+e^{\ts -\mu(x-x_w)}}.
\end{equation}
In the calculation we choose $\mu = 100$, $x_w = 0.05$, and $V_0=1000$.
Initially the free wave function (\ref{eq:4.06}) (with $n_x=n_y=0$ and $\alpha_x=\alpha_y =12.5\sqrt{2}$) impinges on the slit and diffracts.  The parameters of the single-slit calculation are given in Table~\ref{table:4}.
\begin{table}[h]
\caption{Parameters used for the calculations of the wave packet passing through a single slit as shown in Figs.~\ref{fig:07} and \ref{fig:05}.}
\begin{tabular}{l|l}
\hline\hline
$\hbar = 2m = 1$, \ \ $n_x= n_y=0$  & $k_x = 32, k_y =0$ \\
$x_0 = y_0 = -2, \ \ x_\cJ=y_\cM=2$ & $\alpha_x=\alpha_y$ \\
$N = 79, \ \ dt = 10^{-4}$ & $A_x=-0.25, A_y =0$  \\
$\cJ=\cM=200, \ \ r =14, \ \ M = 6$ & $\alpha_x = 12.5\sqrt{2}$ \\
\hline\hline
\end{tabular}
\label{table:4} 
\end{table}
   
\begin{figure}[h]
\resizebox{3.5in}{!}{\includegraphics[angle=-0]
{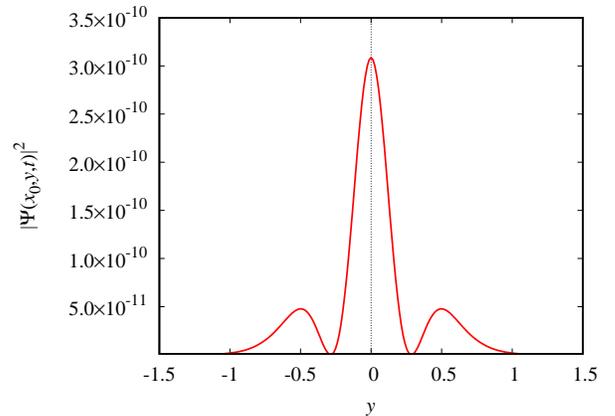}} 
\caption{(Colour online)  The probability density of the wave packet diffracted by a single slit on the plane where $x=x_0=1.25333$ at a time $t=0.0057$.}
\label{fig:07}
\end{figure}

We plot the probability density $|\Psi(x_0,y,t_0)|^2$ as a function of $y$ when $x_0=1.253333$ and $t_0=0.0057$ in Fig.~\ref{fig:07}.  The graph has a remarkable similarity to the Fraunhofer diffraction intensity.  The slit is not of uniform width and hence we cannot compare parameters.  The calculation does indicate that one can study different slit configurations and shapes~\cite{endoh92,*endoh99} using this method.

\begin{figure*}[t]
$\begin{array}{cc}
\vspace{-1cm}
\resizebox{3.5in}{!}{\includegraphics[angle=-0]{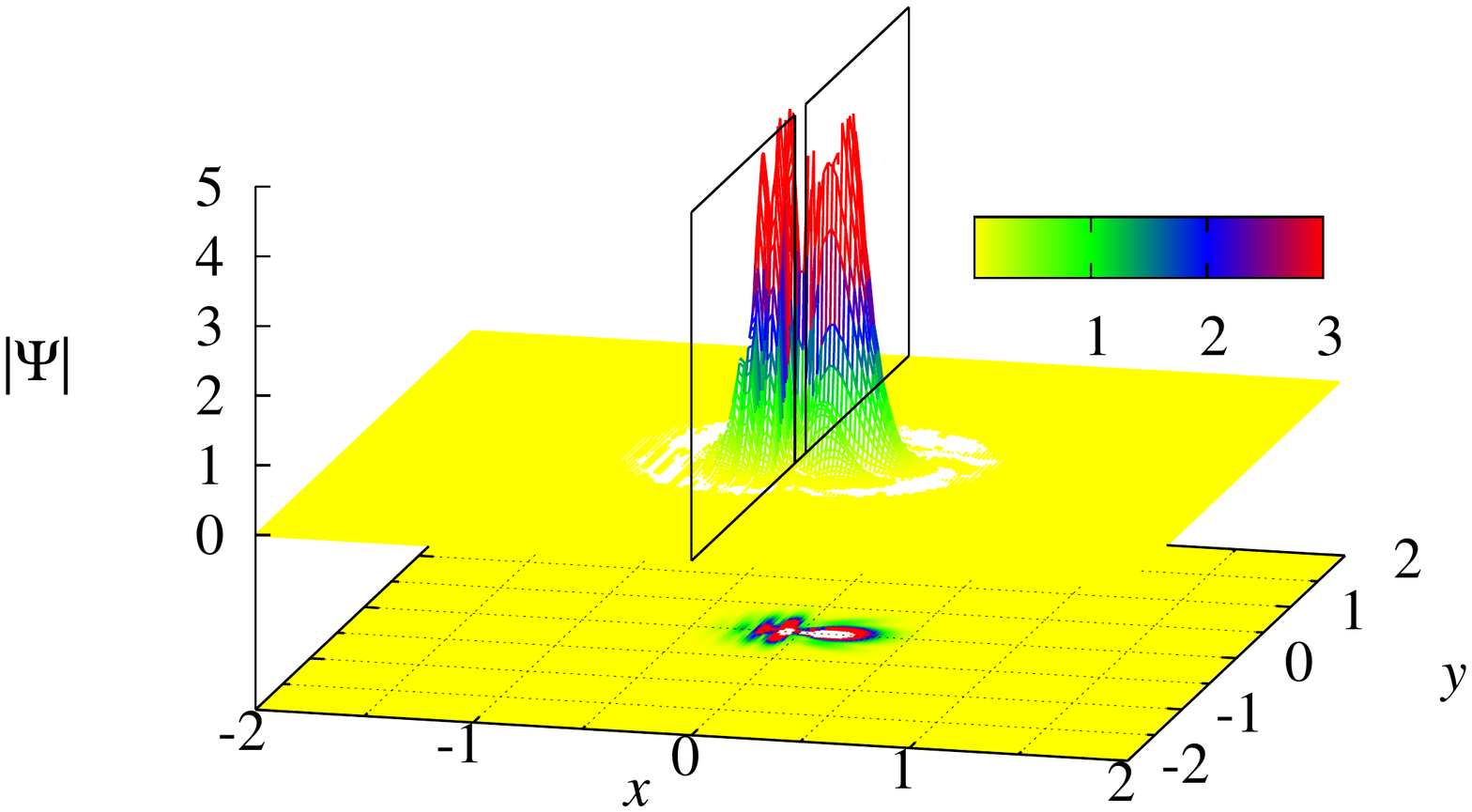}} &
\resizebox{3.5in}{!}{\includegraphics[angle=-0]{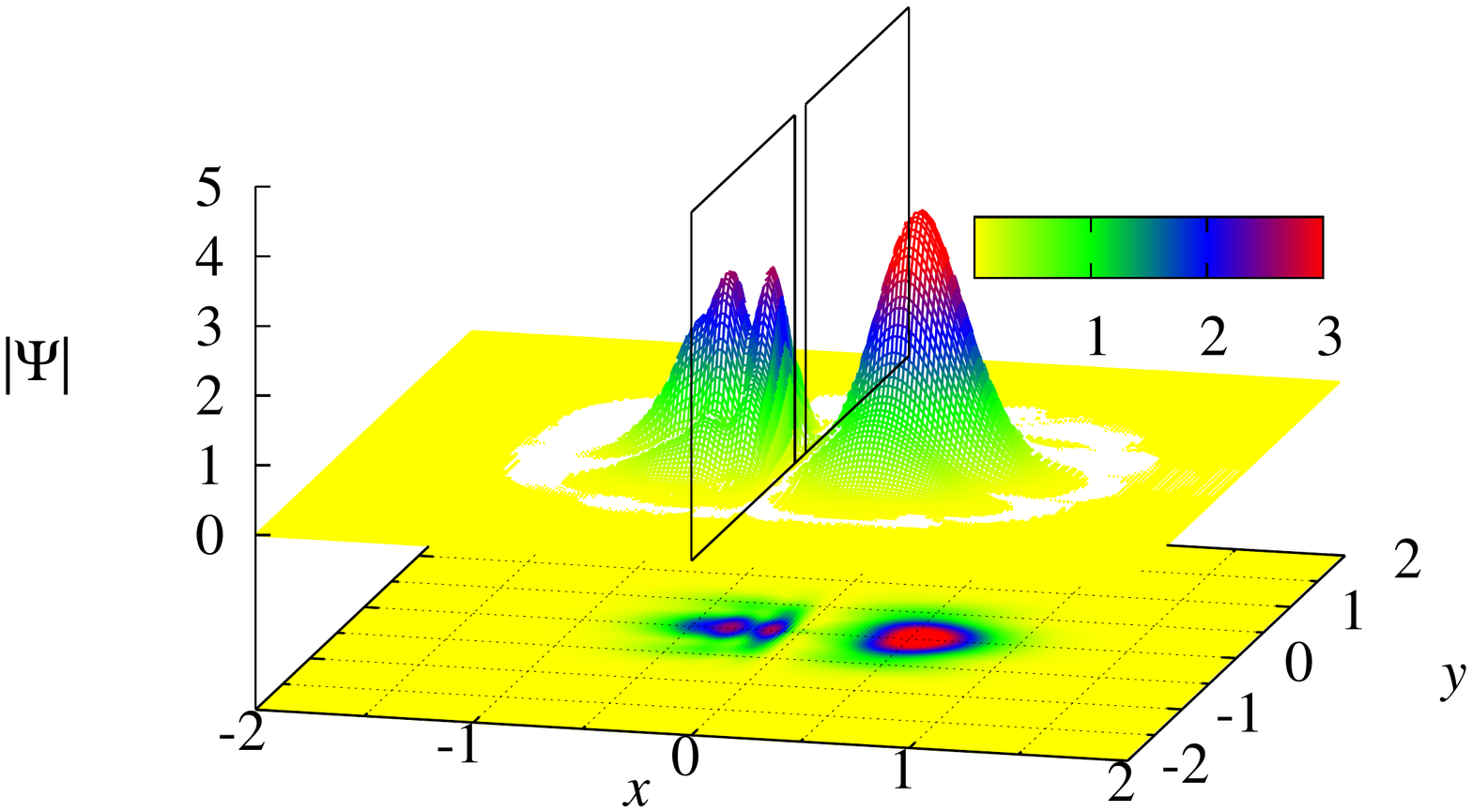}} \\ 
\end{array}$
\caption{The diffraction of a wave packet passing through a single slit, indicated by framed outlines on the graphs.  The time in the left panel is $t = 0.0040$ and in right panel is $t=0.0079$.  See animation~\href{http://physwww.physics.mcmaster.ca/~vandijk/fig:05.mp4}{here}.
}
\label{fig:05}
\end{figure*}

In Fig.~\ref{fig:05} we plot two snapshots of the wave packet passing through the slit.  Since most of the packet is reflected, we multiply the amplitude of the diffracted packet in the figure by ten in order to make the packet's shape in the region beyond the slit more visible.

\section{Discussion}\label{sec:5}

We have shown that the generalized Crank-Nicolson method combined with the alternating-direction implicit procedure is a practical approach to the determination of numerical solutions of the two-dimensional Schr\"odinger equation.  The method allows one to study the efficiency and accuracy systematically as a functions of powers of the temporal and spatial step sizes.  Equation~(\ref{eq:2.15}) is basic to the solution. It can be solved in different ways, but we choose to use an iterative approach which means one must find solutions of linear systems of equations whose coefficients form  banded diagonal matrices.  Since the number of iterations is low, the alternative noniterative approach leads to less sparse matrices and a correspondingly less efficient procedure.

A number of authors~\cite{tian10,gao11,xu12} have considered alternating direction implicit compact finite difference schemes which give errors of order $O(h^5+\Delta t^3)$.  Although they include nonlinear equations in their analysis, they discuss, among others, example 1 of this paper as a test case.  Our scheme gives errors $O(h^{2r} + \Delta t^{2M+1})$ where $M$ and $r$ are positive integers.

The examples demonstrate that this method is capable of accurate solutions even when there is significant fluctuation of the wave function.  The methods described in this paper allow one to obtain a realistic theoretical analysis of the diffraction experiments that have been done recently.  Given the recent attention to the two-dimensional nonlinear Schr\"odinger equation and the Pitaevskii equation, a future project is to expand the method of this paper to such systems, as well as those with time-dependent interactions or source terms.  This in effect is a generalization of the work done earlier on one-dimensional systems~\cite{vandijk14}.

Furthermore it remains to systematically investigate the relative efficiency and accuracy of the approach of this paper to other methods that have been used or proposed.  A generalization to three or higher spatial dimensions and the introduction of transparent boundary conditions are further natural extensions of this work.

\acknowledgments

We are grateful to the Natural Sciences and Engineering Research Council of Canada for the Undergraduate Student Research Awards given to S.-J.P. (2013) and T.V. (2016).


\begin{thebibliography}{36}%
\makeatletter
\providecommand \@ifxundefined [1]{%
 \@ifx{#1\undefined}
}%
\providecommand \@ifnum [1]{%
 \ifnum #1\expandafter \@firstoftwo
 \else \expandafter \@secondoftwo
 \fi
}%
\providecommand \@ifx [1]{%
 \ifx #1\expandafter \@firstoftwo
 \else \expandafter \@secondoftwo
 \fi
}%
\providecommand \natexlab [1]{#1}%
\providecommand \enquote  [1]{``#1''}%
\providecommand \bibnamefont  [1]{#1}%
\providecommand \bibfnamefont [1]{#1}%
\providecommand \citenamefont [1]{#1}%
\providecommand \href@noop [0]{\@secondoftwo}%
\providecommand \href [0]{\begingroup \@sanitize@url \@href}%
\providecommand \@href[1]{\@@startlink{#1}\@@href}%
\providecommand \@@href[1]{\endgroup#1\@@endlink}%
\providecommand \@sanitize@url [0]{\catcode `\\12\catcode `\$12\catcode
  `\&12\catcode `\#12\catcode `\^12\catcode `\_12\catcode `\%12\relax}%
\providecommand \@@startlink[1]{}%
\providecommand \@@endlink[0]{}%
\providecommand \url  [0]{\begingroup\@sanitize@url \@url }%
\providecommand \@url [1]{\endgroup\@href {#1}{\urlprefix }}%
\providecommand \urlprefix  [0]{URL }%
\providecommand \Eprint [0]{\href }%
\providecommand \doibase [0]{http://dx.doi.org/}%
\providecommand \selectlanguage [0]{\@gobble}%
\providecommand \bibinfo  [0]{\@secondoftwo}%
\providecommand \bibfield  [0]{\@secondoftwo}%
\providecommand \translation [1]{[#1]}%
\providecommand \BibitemOpen [0]{}%
\providecommand \bibitemStop [0]{}%
\providecommand \bibitemNoStop [0]{.\EOS\space}%
\providecommand \EOS [0]{\spacefactor3000\relax}%
\providecommand \BibitemShut  [1]{\csname bibitem#1\endcsname}%
\let\auto@bib@innerbib\@empty
\bibitem [{\citenamefont {Tal-Ezer}\ and\ \citenamefont
  {Kosloff}(1984)}]{talezer84}%
  \BibitemOpen
  \bibfield  {author} {\bibinfo {author} {\bibfnamefont {H.}~\bibnamefont
  {Tal-Ezer}}\ and\ \bibinfo {author} {\bibfnamefont {R.}~\bibnamefont
  {Kosloff}},\ }\href@noop {} {\bibfield  {journal} {\bibinfo  {journal} {J.
  Chem. Phys.}\ }\textbf {\bibinfo {volume} {81}},\ \bibinfo {pages} {3967}
  (\bibinfo {year} {1984})}\BibitemShut {NoStop}%
\bibitem [{\citenamefont {Leforestier}\ \emph {et~al.}(1991)\citenamefont
  {Leforestier}, \citenamefont {Bisseling}, \citenamefont {Cerjan},
  \citenamefont {Feit}, \citenamefont {Friesner}, \citenamefont {Guldberg},
  \citenamefont {Hammerich}, \citenamefont {Jolicard}, \citenamefont
  {Karrlein}, \citenamefont {Meyer}, \citenamefont {Lipkin}, \citenamefont
  {Roncero},\ and\ \citenamefont {Kosloff}}]{leforestier91}%
  \BibitemOpen
  \bibfield  {author} {\bibinfo {author} {\bibfnamefont {C.}~\bibnamefont
  {Leforestier}}, \bibinfo {author} {\bibfnamefont {R.~H.}\ \bibnamefont
  {Bisseling}}, \bibinfo {author} {\bibfnamefont {C.}~\bibnamefont {Cerjan}},
  \bibinfo {author} {\bibfnamefont {M.~D.}\ \bibnamefont {Feit}}, \bibinfo
  {author} {\bibfnamefont {R.}~\bibnamefont {Friesner}}, \bibinfo {author}
  {\bibfnamefont {A.}~\bibnamefont {Guldberg}}, \bibinfo {author}
  {\bibfnamefont {A.}~\bibnamefont {Hammerich}}, \bibinfo {author}
  {\bibfnamefont {G.}~\bibnamefont {Jolicard}}, \bibinfo {author}
  {\bibfnamefont {W.}~\bibnamefont {Karrlein}}, \bibinfo {author}
  {\bibfnamefont {H.-D.}\ \bibnamefont {Meyer}}, \bibinfo {author}
  {\bibfnamefont {N.}~\bibnamefont {Lipkin}}, \bibinfo {author} {\bibfnamefont
  {O.}~\bibnamefont {Roncero}}, \ and\ \bibinfo {author} {\bibfnamefont
  {R.}~\bibnamefont {Kosloff}},\ }\href@noop {} {\bibfield  {journal} {\bibinfo
   {journal} {J. Comp. Phys.}\ }\textbf {\bibinfo {volume} {94}},\ \bibinfo
  {pages} {59} (\bibinfo {year} {1991})}\BibitemShut {NoStop}%
\bibitem [{\citenamefont {Puzynin}\ \emph {et~al.}(1999)\citenamefont
  {Puzynin}, \citenamefont {Selin},\ and\ \citenamefont
  {Vinitsky}}]{puzynin99}%
  \BibitemOpen
  \bibfield  {author} {\bibinfo {author} {\bibfnamefont {I.}~\bibnamefont
  {Puzynin}}, \bibinfo {author} {\bibfnamefont {A.}~\bibnamefont {Selin}}, \
  and\ \bibinfo {author} {\bibfnamefont {S.}~\bibnamefont {Vinitsky}},\
  }\href@noop {} {\bibfield  {journal} {\bibinfo  {journal} {Comp. Phys.
  Comm.}\ }\textbf {\bibinfo {volume} {123}},\ \bibinfo {pages} {1} (\bibinfo
  {year} {1999})}\BibitemShut {NoStop}%
\bibitem [{\citenamefont {Puzynin}\ \emph {et~al.}(2000)\citenamefont
  {Puzynin}, \citenamefont {Selin},\ and\ \citenamefont
  {Vinitsky}}]{puzynin00}%
  \BibitemOpen
  \bibfield  {author} {\bibinfo {author} {\bibfnamefont {I.}~\bibnamefont
  {Puzynin}}, \bibinfo {author} {\bibfnamefont {A.}~\bibnamefont {Selin}}, \
  and\ \bibinfo {author} {\bibfnamefont {S.}~\bibnamefont {Vinitsky}},\
  }\href@noop {} {\bibfield  {journal} {\bibinfo  {journal} {Comp. Phys.
  Comm.}\ }\textbf {\bibinfo {volume} {126}},\ \bibinfo {pages} {158} (\bibinfo
  {year} {2000})}\BibitemShut {NoStop}%
\bibitem [{\citenamefont {van Dijk}\ and\ \citenamefont
  {Toyama}(2007)}]{vandijk07}%
  \BibitemOpen
  \bibfield  {author} {\bibinfo {author} {\bibfnamefont {W.}~\bibnamefont {van
  Dijk}}\ and\ \bibinfo {author} {\bibfnamefont {F.~M.}\ \bibnamefont
  {Toyama}},\ }\href@noop {} {\bibfield  {journal} {\bibinfo  {journal} {Phys.
  Rev. E}\ }\textbf {\bibinfo {volume} {75}},\ \bibinfo {pages} {036707 } (\bibinfo {year} {2007})}\BibitemShut {NoStop}%
\bibitem [{\citenamefont {van Dijk}\ \emph {et~al.}(2011)\citenamefont {van
  Dijk}, \citenamefont {Brown},\ and\ \citenamefont {Spyksma}}]{vandijk11}%
  \BibitemOpen
  \bibfield  {author} {\bibinfo {author} {\bibfnamefont {W.}~\bibnamefont {van
  Dijk}}, \bibinfo {author} {\bibfnamefont {J.}~\bibnamefont {Brown}}, \ and\
  \bibinfo {author} {\bibfnamefont {K.}~\bibnamefont {Spyksma}},\ }\href@noop
  {} {\bibfield  {journal} {\bibinfo  {journal} {Phys. Rev. E}\ }\textbf
  {\bibinfo {volume} {84}},\ \bibinfo {pages} {056703} (\bibinfo
  {year} {2011})}\BibitemShut {NoStop}%
\bibitem [{\citenamefont {van Dijk}\ and\ \citenamefont
  {Toyama}(2014)}]{vandijk14a}%
  \BibitemOpen
  \bibfield  {author} {\bibinfo {author} {\bibfnamefont {W.}~\bibnamefont {van
  Dijk}}\ and\ \bibinfo {author} {\bibfnamefont {F.~M.}\ \bibnamefont
  {Toyama}},\ }\href@noop {} {\bibfield  {journal} {\bibinfo  {journal} {Phys.
  Rev. E}\ }\textbf {\bibinfo {volume} {90}},\ \bibinfo {pages} {063309 } (\bibinfo {year} {2014})}\BibitemShut {NoStop}%
\bibitem [{\citenamefont {{van Dijk}}(2016)}]{vandijk16}%
  \BibitemOpen
  \bibfield  {author} {\bibinfo {author} {\bibfnamefont {W.}~\bibnamefont {{van
  Dijk}}},\ }\href@noop {} {\bibfield  {journal} {\bibinfo  {journal} {Phys.
  Rev. E}\ }\textbf {\bibinfo {volume} {93}},\ \bibinfo {pages} {063307 } (\bibinfo {year} {2016})}\BibitemShut {NoStop}%
\bibitem [{\citenamefont {Form\'anek}\ \emph {et~al.}(2010)\citenamefont
  {Form\'anek}, \citenamefont {V\'a{\v n}a},\ and\ \citenamefont
  {Houfek}}]{formanek10}%
  \BibitemOpen
  \bibfield  {author} {\bibinfo {author} {\bibfnamefont {M.}~\bibnamefont
  {Form\'anek}}, \bibinfo {author} {\bibfnamefont {M.}~\bibnamefont {V\'a{\v
  n}a}}, \ and\ \bibinfo {author} {\bibfnamefont {K.}~\bibnamefont {Houfek}},\
  }in\ \href@noop {} {\emph {\bibinfo {booktitle} {Numerical Analysis and
  Applied Mathematics, International Conference 2010}}},\ \bibinfo {editor}
  {edited by\ \bibinfo {editor} {\bibfnamefont {T.~E.}\ \bibnamefont {Simos}},
  \bibinfo {editor} {\bibfnamefont {G.}~\bibnamefont {Psihoyios}}, \ and\
  \bibinfo {editor} {\bibfnamefont {C.}~\bibnamefont {Tsitouras}}}\ (\bibinfo
  {publisher} {American Institute of Physics},\ \bibinfo {year} {2010})\ pp.\
  \bibinfo {pages} {667--670}\BibitemShut {NoStop}%
\bibitem [{\citenamefont {Gusev}\ \emph {et~al.}(2014)\citenamefont {Gusev},
  \citenamefont {Chuluunbaatar}, \citenamefont {Vinitsky},\ and\ \citenamefont
  {Abrashevich}}]{gusev14}%
  \BibitemOpen
  \bibfield  {author} {\bibinfo {author} {\bibfnamefont {A.~A.}\ \bibnamefont
  {Gusev}}, \bibinfo {author} {\bibfnamefont {O.}~\bibnamefont
  {Chuluunbaatar}}, \bibinfo {author} {\bibfnamefont {S.~I.}\ \bibnamefont
  {Vinitsky}}, \ and\ \bibinfo {author} {\bibfnamefont {A.~G.}\ \bibnamefont
  {Abrashevich}},\ }\href@noop {} {\bibfield  {journal} {\bibinfo  {journal}
  {Math. Mod. and Geom.}\ }\textbf {\bibinfo {volume} {2}},\ \bibinfo {pages}
  {33} (\bibinfo {year} {2014})}\BibitemShut {NoStop}  %
\bibitem [{\citenamefont {Chuluunbaatar}\ \emph
  {et~al.}(2008{\natexlab{a}})\citenamefont {Chuluunbaatar}, \citenamefont  
  {Derbov}, \citenamefont {Galtbayar}, \citenamefont {Gusev}, \citenamefont
  {Kashiev}, \citenamefont {Vinitsky},\ and\ \citenamefont
  {Zhanlav}}]{chuluunbaatar08}%
  \BibitemOpen 
  \bibfield  {author} {\bibinfo {author} {\bibfnamefont {A.}~\bibnamefont
  {Chuluunbaatar}}, \bibinfo {author} {\bibfnamefont {V.~L.}\ \bibnamefont
  {Derbov}}, \bibinfo {author} {\bibfnamefont {A.}~\bibnamefont {Galtbayar}},
  \bibinfo {author} {\bibfnamefont {A.~A.}\ \bibnamefont {Gusev}}, \bibinfo
  {author} {\bibfnamefont {M.~S.}\ \bibnamefont {Kashiev}}, \bibinfo {author}
  {\bibfnamefont {S.~I.}\ \bibnamefont {Vinitsky}}, \ and\ \bibinfo {author}
  {\bibfnamefont {T.}~\bibnamefont {Zhanlav}},\ }\href@noop {} {\bibfield
  {journal} {\bibinfo  {journal} {J. Phys. A: Math. Theor.}\ }\textbf {\bibinfo
  {volume} {41}},\ \bibinfo {pages} {295203} (\bibinfo {year}
  {2008}{\natexlab{a}})}\BibitemShut {NoStop}  %
\bibitem [{\citenamefont {Chuluunbaatar}\ \emph
  {et~al.}(2008{\natexlab{b}})\citenamefont {Chuluunbaatar}, \citenamefont
  {Gusev}, \citenamefont {Vinitsky}, \citenamefont {Derbov}, \citenamefont
  {Galtbayar},\ and\ \citenamefont {Zhanlav}}]{chuluunbaatar08a}%
  \BibitemOpen
  \bibfield  {author} {\bibinfo {author} {\bibfnamefont {O.}~\bibnamefont
  {Chuluunbaatar}}, \bibinfo {author} {\bibfnamefont {A.~A.}\ \bibnamefont
  {Gusev}}, \bibinfo {author} {\bibfnamefont {S.~I.}\ \bibnamefont {Vinitsky}},
  \bibinfo {author} {\bibfnamefont {V.~L.}\ \bibnamefont {Derbov}}, \bibinfo
  {author} {\bibfnamefont {A.}~\bibnamefont {Galtbayar}}, \ and\ \bibinfo
  {author} {\bibfnamefont {T.}~\bibnamefont {Zhanlav}},\ }\href@noop {}
  {\bibfield  {journal} {\bibinfo  {journal} {Phys. Rev. E}\ }\textbf {\bibinfo
  {volume} {78}},\ \bibinfo {pages} {017701} (\bibinfo {year}
  {2008}{\natexlab{b}})}\BibitemShut {NoStop}  %
\bibitem [{\citenamefont {Crank}\ and\ \citenamefont
  {Nicolson}(1947)}]{crank47}%
  \BibitemOpen
  \bibfield  {author} {\bibinfo {author} {\bibfnamefont {J.}~\bibnamefont
  {Crank}}\ and\ \bibinfo {author} {\bibfnamefont {E.}~\bibnamefont
  {Nicolson}},\ }\href@noop {} {\bibfield  {journal} {\bibinfo  {journal}
  {Proc. Camb. Phil. Soc.}\ }\textbf {\bibinfo {volume} {43}},\ \bibinfo
  {pages} {50} (\bibinfo {year} {1947})},\ \bibinfo {note} {reprinted in
  Advances in Computational Mathematics \textbf{6}, 207 (1996)}\BibitemShut {NoStop}%
\bibitem [{\citenamefont {Feit}\ \emph {et~al.}(1982)\citenamefont {Feit},
  \citenamefont {Fleck},\ and\ \citenamefont {Steiger}}]{feit82}%
  \BibitemOpen
  \bibfield  {author} {\bibinfo {author} {\bibfnamefont {M.~D.}\ \bibnamefont
  {Feit}}, \bibinfo {author} {\bibfnamefont {J.~A.}\ \bibnamefont {Fleck}}, \
  and\ \bibinfo {author} {\bibfnamefont {A.}~\bibnamefont {Steiger}},\
  }\href@noop {} {\bibfield  {journal} {\bibinfo  {journal} {J. Comp. Phys.}\
  }\textbf {\bibinfo {volume} {47}},\ \bibinfo {pages} {412} (\bibinfo {year}
  {1982})}\BibitemShut {NoStop}%
\bibitem [{\citenamefont {Park}\ and\ \citenamefont {Light}(1986)}]{park86}%
  \BibitemOpen
  \bibfield  {author} {\bibinfo {author} {\bibfnamefont {T.~J.}\ \bibnamefont
  {Park}}\ and\ \bibinfo {author} {\bibfnamefont {J.~C.}\ \bibnamefont
  {Light}},\ }\href@noop {} {\bibfield  {journal} {\bibinfo  {journal} {J.
  Chem. Phys.}\ }\textbf {\bibinfo {volume} {85}},\ \bibinfo {pages} {5870}
  (\bibinfo {year} {1986})}\BibitemShut {NoStop}%
\bibitem [{\citenamefont {Tian}\ and\ \citenamefont {Yu}(2010)}]{tian10}%
  \BibitemOpen
  \bibfield  {author} {\bibinfo {author} {\bibfnamefont {Z.~F.}\ \bibnamefont
  {Tian}}\ and\ \bibinfo {author} {\bibfnamefont {P.~X.}\ \bibnamefont {Yu}},\
  }\href@noop {} {\bibfield  {journal} {\bibinfo  {journal} {Comp. Phys.
  Comm.}\ }\textbf {\bibinfo {volume} {181}},\ \bibinfo {pages} {861} (\bibinfo
  {year} {2010})}\BibitemShut {NoStop}%
\bibitem [{\citenamefont {Xu}\ and\ \citenamefont {Zhang}(2012)}]{xu12}%
  \BibitemOpen
  \bibfield  {author} {\bibinfo {author} {\bibfnamefont {Y.}~\bibnamefont
  {Xu}}\ and\ \bibinfo {author} {\bibfnamefont {L.}~\bibnamefont {Zhang}},\
  }\href@noop {} {\bibfield  {journal} {\bibinfo  {journal} {Comp. Phys.
  Comm.}\ }\textbf {\bibinfo {volume} {183}},\ \bibinfo {pages} {1082}
  (\bibinfo {year} {2012})}\BibitemShut {NoStop}%
\bibitem [{\citenamefont {Gao}\ and\ \citenamefont {Mei}(2016)}]{gao16}%
  \BibitemOpen
  \bibfield  {author} {\bibinfo {author} {\bibfnamefont {Y.}~\bibnamefont
  {Gao}}\ and\ \bibinfo {author} {\bibfnamefont {L.}~\bibnamefont {Mei}},\
  }\href@noop {} {\bibfield  {journal} {\bibinfo  {journal} {Appl. Num. Math.}\
  }\textbf {\bibinfo {volume} {109}},\ \bibinfo {pages} {41} (\bibinfo {year}
  {2016})}\BibitemShut {NoStop}%
\bibitem [{\citenamefont {Symes}\ \emph {et~al.}(2016)\citenamefont {Symes},
  \citenamefont {McLachlan},\ and\ \citenamefont {Blakie}}]{symes16}%
  \BibitemOpen
  \bibfield  {author} {\bibinfo {author} {\bibfnamefont {L.~M.}\ \bibnamefont
  {Symes}}, \bibinfo {author} {\bibfnamefont {R.~I.}\ \bibnamefont
  {McLachlan}}, \ and\ \bibinfo {author} {\bibfnamefont {P.~B.}\ \bibnamefont
  {Blakie}},\ }\href@noop {} {\bibfield  {journal} {\bibinfo  {journal} {Phys.
  Rev. E}\ }\textbf {\bibinfo {volume} {93}},\ \bibinfo {pages} {053309 } (\bibinfo {year} {2016})}\BibitemShut {NoStop}%
\bibitem [{\citenamefont {Wang}\ \emph {et~al.}(2016)\citenamefont {Wang},
  \citenamefont {Huang}, \citenamefont {Tian},\ and\ \citenamefont
  {Zhou}}]{wang16}%
  \BibitemOpen
  \bibfield  {author} {\bibinfo {author} {\bibfnamefont {J.}~\bibnamefont
  {Wang}}, \bibinfo {author} {\bibfnamefont {Y.}~\bibnamefont {Huang}},
  \bibinfo {author} {\bibfnamefont {Z.}~\bibnamefont {Tian}}, \ and\ \bibinfo
  {author} {\bibfnamefont {J.}~\bibnamefont {Zhou}},\ }\href@noop {} {\bibfield
   {journal} {\bibinfo  {journal} {Computers \& Math. with Appl.}\ }\textbf
  {\bibinfo {volume} {71}},\ \bibinfo {pages} {1960} (\bibinfo {year}
  {2016})}\BibitemShut {NoStop}%
\bibitem [{\citenamefont {Zhang}\ and\ \citenamefont {Chen}(2016)}]{zhang16}%
  \BibitemOpen
  \bibfield  {author} {\bibinfo {author} {\bibfnamefont {S.}~\bibnamefont
  {Zhang}}\ and\ \bibinfo {author} {\bibfnamefont {S.}~\bibnamefont {Chen}},\
  }\href@noop {} {\bibfield  {journal} {\bibinfo  {journal} {Computers \& Math.
  with Appl.}\ }\textbf {\bibinfo {volume} {72}},\ \bibinfo {pages} {2143}
  (\bibinfo {year} {2016})}\BibitemShut {NoStop}%
\bibitem [{\citenamefont {Gaspar}\ \emph {et~al.}(2014)\citenamefont {Gaspar},
  \citenamefont {Rodrigo}, \citenamefont {\u{C}iegis},\ and\ \citenamefont
  {Mirinavi\u{c}ius}}]{gaspar14}%
  \BibitemOpen
  \bibfield  {author} {\bibinfo {author} {\bibfnamefont {F.~J.}\ \bibnamefont
  {Gaspar}}, \bibinfo {author} {\bibfnamefont {C.}~\bibnamefont {Rodrigo}},
  \bibinfo {author} {\bibfnamefont {R.}~\bibnamefont {\u{C}iegis}}, \ and\
  \bibinfo {author} {\bibfnamefont {A.}~\bibnamefont {Mirinavi\u{c}ius}},\
  }\href@noop {} {\bibfield  {journal} {\bibinfo  {journal} {Int. J. Numer.
  Anal. \& Modeling}\ }\textbf {\bibinfo {volume} {1}},\ \bibinfo {pages} {131}
  (\bibinfo {year} {2014})}\BibitemShut {NoStop}%
\bibitem [{\citenamefont {Peaceman}\ and\ \citenamefont
  {H.~H.~Rachford}(1955)}]{peaceman55}%
  \BibitemOpen
  \bibfield  {author} {\bibinfo {author} {\bibfnamefont {D.~W.}\ \bibnamefont
  {Peaceman}}\ and\ \bibinfo {author} {\bibfnamefont {J.}~\bibnamefont
  {H.~H.~Rachford}},\ }\href@noop {} {\bibfield  {journal} {\bibinfo  {journal}
  {J. Soc. Indust. Appl. Math.}\ }\textbf {\bibinfo {volume} {3}},\ \bibinfo
  {pages} {28} (\bibinfo {year} {1955})}\BibitemShut {NoStop}%
\bibitem [{\citenamefont {Cerjan}\ and\ \citenamefont
  {Kulander}(1991)}]{cerjan91}%
  \BibitemOpen
  \bibfield  {author} {\bibinfo {author} {\bibfnamefont {C.}~\bibnamefont
  {Cerjan}}\ and\ \bibinfo {author} {\bibfnamefont {K.~C.}\ \bibnamefont
  {Kulander}},\ }\href@noop {} {\bibfield  {journal} {\bibinfo  {journal}
  {Comput. Phys. Commun.}\ }\textbf {\bibinfo {volume} {63}},\ \bibinfo {pages}
  {529 } (\bibinfo {year} {1991})}\BibitemShut {NoStop}%
\bibitem [{\citenamefont {Wang}\ and\ \citenamefont {Shao}(2009)}]{wang09}%
  \BibitemOpen
  \bibfield  {author} {\bibinfo {author} {\bibfnamefont {Z.}~\bibnamefont
  {Wang}}\ and\ \bibinfo {author} {\bibfnamefont {H.}~\bibnamefont {Shao}},\
  }\href@noop {} {\bibfield  {journal} {\bibinfo  {journal} {Comp. Phys.
  Comm.}\ }\textbf {\bibinfo {volume} {180}},\ \bibinfo {pages} {842} (\bibinfo
  {year} {2009})}\BibitemShut {NoStop}%
\bibitem [{\citenamefont {Shao}\ and\ \citenamefont {Wang}(2009)}]{shao09}%
  \BibitemOpen
  \bibfield  {author} {\bibinfo {author} {\bibfnamefont {H.}~\bibnamefont
  {Shao}}\ and\ \bibinfo {author} {\bibfnamefont {Z.}~\bibnamefont {Wang}},\
  }\href {\doibase 10.1103/PhysRevE.79.056705} {\bibfield  {journal} {\bibinfo
  {journal} {Phys. Rev. E}\ }\textbf {\bibinfo {volume} {79}},\ \bibinfo
  {pages} {056705} (\bibinfo {year} {2009})}\BibitemShut {NoStop}%
\bibitem [{\citenamefont {Peters}\ and\ \citenamefont
  {Maley}(1968)}]{peters68}%
  \BibitemOpen
  \bibfield  {author} {\bibinfo {author} {\bibfnamefont {G.~O.}\ \bibnamefont
  {Peters}}\ and\ \bibinfo {author} {\bibfnamefont {C.~E.}\ \bibnamefont
  {Maley}},\ }\href@noop {} {\bibfield  {journal} {\bibinfo  {journal} {Am.
  Math. Monthly}\ }\textbf {\bibinfo {volume} {75}},\ \bibinfo {pages} {741}
  (\bibinfo {year} {1968})}\BibitemShut {NoStop}%
\bibitem [{\citenamefont {$\breve{\rm C}$iegis}\ \emph
  {et~al.}(2013)\citenamefont {$\breve{\rm C}$iegis}, \citenamefont
  {Mirinavi$\breve{\rm c}$ius},\ and\ \citenamefont {Radziunas}}]{ciegis13}%
  \BibitemOpen
  \bibfield  {author} {\bibinfo {author} {\bibfnamefont {R.}~\bibnamefont
  {$\breve{\rm C}$iegis}}, \bibinfo {author} {\bibfnamefont {A.}~\bibnamefont
  {Mirinavi$\breve{\rm c}$ius}}, \ and\ \bibinfo {author} {\bibfnamefont
  {M.}~\bibnamefont {Radziunas}},\ }\href@noop {} {\bibfield  {journal}
  {\bibinfo  {journal} {Comp. Meth. Appl. Math.}\ }\textbf {\bibinfo {volume}
  {13}},\ \bibinfo {pages} {237} (\bibinfo {year} {2013})}\BibitemShut
  {NoStop}%
\bibitem [{\citenamefont {van Dijk}\ \emph {et~al.}(2014)\citenamefont {van
  Dijk}, \citenamefont {Toyama}, \citenamefont {Prins},\ and\ \citenamefont
  {Spyksma}}]{vandijk14}%
  \BibitemOpen
  \bibfield  {author} {\bibinfo {author} {\bibfnamefont {W.}~\bibnamefont {van
  Dijk}}, \bibinfo {author} {\bibfnamefont {F.~M.}\ \bibnamefont {Toyama}},
  \bibinfo {author} {\bibfnamefont {S.~J.}\ \bibnamefont {Prins}}, \ and\
  \bibinfo {author} {\bibfnamefont {K.}~\bibnamefont {Spyksma}},\ }\href@noop
  {} {\bibfield  {journal} {\bibinfo  {journal} {Am. J. Phys.}\ }\textbf
  {\bibinfo {volume} {82}},\ \bibinfo {pages} {955} (\bibinfo {year}
  {2014})}\BibitemShut {NoStop}%
\bibitem [{\citenamefont {Abramowitz}\ and\ \citenamefont
  {Stegun}(1965)}]{abramowitz65}%
  \BibitemOpen
  \bibfield  {author} {\bibinfo {author} {\bibfnamefont {M.}~\bibnamefont
  {Abramowitz}}\ and\ \bibinfo {author} {\bibfnamefont {I.~A.}\ \bibnamefont
  {Stegun}},\ }\href@noop {} {\emph {\bibinfo {title} {Handbook of Mathematical
  Functions}}}\ (\bibinfo  {publisher} {Dover Publications, Inc.},\ \bibinfo
  {address} {New York},\ \bibinfo {year} {1965})\BibitemShut {NoStop}%
\bibitem [{\citenamefont {Galbraith}\ \emph {et~al.}(1984)\citenamefont
  {Galbraith}, \citenamefont {Ching},\ and\ \citenamefont
  {Abraham}}]{galbraith84}%
  \BibitemOpen
  \bibfield  {author} {\bibinfo {author} {\bibfnamefont {I.}~\bibnamefont
  {Galbraith}}, \bibinfo {author} {\bibfnamefont {Y.~S.}\ \bibnamefont
  {Ching}}, \ and\ \bibinfo {author} {\bibfnamefont {E.}~\bibnamefont
  {Abraham}},\ }\href@noop {} {\bibfield  {journal} {\bibinfo  {journal} {Am.
  J. Phys.}\ }\textbf {\bibinfo {volume} {52}},\ \bibinfo {pages} {60}
  (\bibinfo {year} {1984})}\BibitemShut {NoStop}%
\bibitem [{\citenamefont {Endoh}\ \emph {et~al.}(1992)\citenamefont {Endoh},
  \citenamefont {Sasa},\ and\ \citenamefont {Muto}}]{endoh92}%
  \BibitemOpen
  \bibfield  {author} {\bibinfo {author} {\bibfnamefont {A.}~\bibnamefont
  {Endoh}}, \bibinfo {author} {\bibfnamefont {S.}~\bibnamefont {Sasa}}, \ and\
  \bibinfo {author} {\bibfnamefont {S.}~\bibnamefont {Muto}},\ }\href@noop {}
  {\bibfield  {journal} {\bibinfo  {journal} {Appl. Phys. Lett.}\ }\textbf
  {\bibinfo {volume} {61}},\ \bibinfo {pages} {52} (\bibinfo {year}
  {1992})}\BibitemShut {NoStop}%
\bibitem [{\citenamefont {Endoh}\ \emph {et~al.}(1999)\citenamefont {Endoh},
  \citenamefont {Sasa}, \citenamefont {Arimoto},\ and\ \citenamefont
  {Muto}}]{endoh99}%
  \BibitemOpen
  \bibfield  {author} {\bibinfo {author} {\bibfnamefont {A.}~\bibnamefont
  {Endoh}}, \bibinfo {author} {\bibfnamefont {S.}~\bibnamefont {Sasa}},
  \bibinfo {author} {\bibfnamefont {H.}~\bibnamefont {Arimoto}}, \ and\
  \bibinfo {author} {\bibfnamefont {S.}~\bibnamefont {Muto}},\ }\href@noop {}
  {\bibfield  {journal} {\bibinfo  {journal} {Am. J. Phys.}\ }\textbf {\bibinfo
  {volume} {86}},\ \bibinfo {pages} {6249} (\bibinfo {year}
  {1999})}\BibitemShut {NoStop}%
\bibitem [{\citenamefont {Bach}\ \emph {et~al.}(2013)\citenamefont {Bach},
  \citenamefont {Pope}, \citenamefont {Liou},\ and\ \citenamefont
  {Batelaan}}]{bach13}%
  \BibitemOpen
  \bibfield  {author} {\bibinfo {author} {\bibfnamefont {R.}~\bibnamefont
  {Bach}}, \bibinfo {author} {\bibfnamefont {D.}~\bibnamefont {Pope}}, \bibinfo
  {author} {\bibfnamefont {S.-H.}\ \bibnamefont {Liou}}, \ and\ \bibinfo
  {author} {\bibfnamefont {H.}~\bibnamefont {Batelaan}},\ }\href@noop {}
  {\bibfield  {journal} {\bibinfo  {journal} {New J. Phys.}\ }\textbf {\bibinfo
  {volume} {15}},\ \bibinfo {pages} {033018 (7 pages)} (\bibinfo {year}
  {2013})}\BibitemShut {NoStop}%
\bibitem [{\citenamefont {Khatua}\ \emph {et~al.}(2014)\citenamefont {Khatua},
  \citenamefont {Bansal},\ and\ \citenamefont {Shahar}}]{khatua14}%
  \BibitemOpen
  \bibfield  {author} {\bibinfo {author} {\bibfnamefont {P.}~\bibnamefont
  {Khatua}}, \bibinfo {author} {\bibfnamefont {B.}~\bibnamefont {Bansal}}, \
  and\ \bibinfo {author} {\bibfnamefont {D.}~\bibnamefont {Shahar}},\
  }\href@noop {} {\bibfield  {journal} {\bibinfo  {journal} {Phys. Rev. Lett.}\
  }\textbf {\bibinfo {volume} {112}},\ \bibinfo {pages} {010403 (5 pages)}
  (\bibinfo {year} {2014})}\BibitemShut {NoStop}%
\bibitem [{\citenamefont {Gao}\ and\ \citenamefont {Xie}(2015)}]{gao11}%
  \BibitemOpen
  \bibfield  {author} {\bibinfo {author} {\bibfnamefont {Z.}~\bibnamefont
  {Gao}}\ and\ \bibinfo {author} {\bibfnamefont {S.}~\bibnamefont {Xie}},\
  }\href@noop {} {\bibfield  {journal} {\bibinfo  {journal} {Applied Numerical
  Mathematics}\ }\textbf {\bibinfo {volume} {61}},\ \bibinfo {pages} {593}
  (\bibinfo {year} {2015})}\BibitemShut {NoStop}%
\end{thebibliography}

%

\end{document}